\documentclass[twocolumn]{aastex701}

\usepackage{amsmath}
\usepackage{amssymb}
\usepackage{lipsum}
\usepackage{ragged2e}
\usepackage{booktabs}
\usepackage{caption}
\usepackage[english]{babel}
\PassOptionsToPackage{disable}{longtable} % Ensure longtable is disabled to avoid conflicts
\usepackage{adjustbox} % Optional: for adjusting table size
\usepackage{graphicx}  % For resizing tables
\usepackage{tabularx}
\usepackage{subcaption}

\begin{document}

\title{Constraining the Leptonic and Hadronic Contributions to the Gamma-Ray Emission of SNR HESS~J1731$-$347 with CTAO Forecasts}

\author[orcid=0000-0002-9446-4396, gname=D\'ebora Beatriz, sname='G\¨otz']{D. B. G\"otz}
%%\altaffiliation{Kitt Peak National Observatory}
\affiliation{Programa de Pós-graduação em Física, Universidade Estadual de Londrina, 86057-970, Londrina, PR, Brazil}
\email[show]{debora.beatriz.gotz@uel.br}

\author[0000-0002-5438-3460]{M. F. Sousa}
\affiliation{Instituto de Física de São Carlos, Universidade de São Paulo, Av. Trabalhador São-carlense 400, São Carlos, SP, Brazil}
\email[show]{mfelipe.sousa@ifsc.usp.br}

\author[0000-0002-6463-2272]{R. C. Anjos}
\affiliation{Departamento de Engenharia e Ciências Exatas, Universidade Federal do Paraná, Pioneiro, 2153, 85950-000, Palotina, PR, Brazil}
\affiliation{Centro de Artes, Humanidades e Tecnologia, Universidade Federal de São Carlos, R.\ Dr.\ Eduardo Nielsen, 420, Jardim Congonhas, São José do Rio Preto, 15030-070, SP, Brazil}
\affiliation{Programa de Pós-graduação em Física, Universidade Estadual de Londrina, 86057-970, Londrina, PR, Brazil}
\affiliation{Programa de Pós-Graduação em Astrofísica, Cosmologia e Gravitação, Universidade Federal do Espírito Santo, 29075--910, Vitória, ES, Brazil}
\affiliation{Programa de Pós-graduação em Física e Astronomia, Universidade Tecnológica do Paraná, 80230-901, Curitiba, PR, Brazil}
\affiliation{Programa de Pós-graduação em Física Aplicada, Universidade Federal da Integração Latino-Americana, 85867-670, Foz do Igua\c{c}u, PR, Brazil}
\email[show]{\\ritacassia@ufscar.br}

\begin{abstract}
We present a Bayesian broadband spectral analysis of the supernova
remnant HESS~J1731$-$347, combining archival radio, X-ray, and H.E.S.S.
data with a new analysis of 16~yr of \textit{Fermi}-LAT observations. The
LAT data allow the GeV emission of the western molecular-cloud core
(MC-core) to be separated from that of the remaining shell. We fit purely
leptonic and lepto-hadronic models with and without the MC-core region.
Within a one-zone model with fixed distance and ambient density, the
lepto-hadronic scenario is strongly preferred by the Bayesian Information
Criterion (BIC) in both cases. Excluding the MC-core, we obtain
$\Delta{\rm BIC}=33.67$,a magnetic field of $B = 84.9^{+27.5}_{-22.9}~\mu{\rm G}$, 
a hard proton index of $\alpha_p = 1.69 \pm 0.10$, and a proton energy of
$W_p = (4.37^{+0.64}_{-0.51})\times10^{48}$~erg for $n_{\rm H}=20~{\rm cm^{-3}}$. 
The preference for a hadronic contribution is therefore not driven by the
MC-core emission alone. CTAO simulations show that a 50~h observation with
the southern array would detect the source at $13\sigma$ to $18\sigma$ and substantially improve the constraints on its TeV spectrum.
\end{abstract}

\keywords{\uat{Gamma-ray sources}{633} --- \uat{High-energy astrophysics}{739} --- \uat{Non-thermal radiation sources}{1119} --- \uat{Supernova remnants}{1667}}

% ============================================================
\section{Introduction}
\label{sec:intro}
% ============================================================

The origin of Galactic cosmic rays (CR) remains a central
problem in high-energy astrophysics. Supernova remnants (SNRs)
are widely considered among the primary candidates for accelerating
particles up to the ``knee'' of the CR spectrum, near
$3 \times 10^{15}$~eV, via diffusive shock acceleration (DSA)~\citep[see][and references therein]{Bell1978,2021A&A...650A..62C,2023PPCF...65a4002B}.
Observational support for this scenario includes the detection of non-thermal X-ray synchrotron emission
in several young remnants, indicating the presence of electrons
accelerated to multi-TeV energies~\citep{Koyama1995}. In
addition, GeV and TeV gamma-ray ($\gamma$-ray) observations of sources such
as Cas~A, RCW~86, and RX~J1713.7$-$3946 provide direct
constraints on the high-energy particle populations. However,
disentangling leptonic emission, primarily inverse-Compton (IC)
scattering, from hadronic processes, namely neutral pion decay
following proton-proton interactions, remains a key challenge.
Both scenarios can reproduce similar broadband spectra under
different environmental conditions
\citep{Abdo2010, Ackermann2013}.

Among TeV-emitting SNRs, HESS~J1731$-$347
(G353.6$-$0.7) is one of only a handful of remnants with
a fully resolved shell-type morphology at very high energies (VHEs),
alongside RX~J1713.7$-$3946, RCW~86, SN~1006, and Vela~Jr.
\citep{HESS2018RXJ1713,HESS2018RCW86,HESS2010SN1006,HESS2018VelaJr}.
The source was first reported as an unidentified VHE source by~\citet{Aharonian2008}.
A deep H.E.S.S. exposure of $\sim$59~h
subsequently revealed a nearly circular TeV shell of radius
$r \sim 0.25^{\circ}$, in close spatial correspondence with the
radio shell G353.6$-$0.7 discovered by~\citet{Tian2008}, and
firmly established the association between the two~\citep{HESS2011}. Like the other members of this class,
HESS~J1731$-$347 has a predominantly non-thermal
X-ray spectrum along the entire shell, with no detectable
thermal component~\citep{Doroshenko2017}, a property most
naturally explained by synchrotron emission from multi-TeV
electrons accelerated at the forward shock propagating into a
tenuous medium. The detection of the central compact object
XMMU~J173203.3$-$344518 near the geometrical center of the
remnant~\citep{Halpern2010, Tian2010} confirms a core-collapse
origin and identifies the compact stellar remnant of the
explosion.

Despite this clear morphological picture, the fundamental
parameters of the remnant remain debated, and this uncertainty
propagates directly into any broadband modeling effort. The
distance has been inferred from two independent lines of
evidence that do not fully agree. The smaller value of
$d \simeq 3.2$~kpc rests on the spatial association with the
H\,{\sc II} region G353.42$-$0.37~\citep{Tian2008} and is
reinforced by the joint analysis of X-ray absorption columns
and H\,{\sc I}+CO line-of-sight column densities by~\citet{Maxted2018}, who found that the foreground absorption is
consistent with material up to, but not beyond, the Scutum-Crux
Galactic arm. In contrast, the interpretation of~\citet{Fukuda2014}, based on a kinematic identification of an
H\,{\sc I} cavity along the line of sight, favors a larger
distance of $5.2$ to $6.1$~kpc. Estimates of the age of the
remnant, which depend on both the distance and the assumed
shock dynamics, span a correspondingly wide interval of
$2$ to $27$~kyr~(\citealp{HESS2011,Fukuda2014,Doroshenko2017}).

The ambient medium around the remnant is manifestly non-uniform.
The western shell is spatially coincident with a molecular cloud
at $\sim$3.2~kpc and exhibits noticeably suppressed X-ray
emission relative to the rest of the remnant, an asymmetry that~\citet{Cui2019} interpret as evidence of an ongoing shock-cloud
interaction, in which the decelerated shock can no longer
sustain efficient electron acceleration or strong magnetic
turbulence. The Mopra CO/CS survey of~\citet{Maxted2018} further
shows that, while dense molecular gas is found to the north of
the remnant, coincident with G353.42$-$0.37 and with the
neighboring TeV source HESS~J1729$-$345, the bulk of the shell
is not overlaid by dense material, a picture consistent with a
progenitor wind-blown cavity scenario.

At GeV energies, early analyses with the Large Area Telescope on board
the \textit{Fermi} Gamma-ray Space Telescope (\textit{Fermi}-LAT) provided only upper limits
toward HESS~J1731$-$347 \citep{Yang2014}. Subsequent analyses based on
Pass~8 data established the GeV counterpart of the remnant.
\citet{Condon_2017} reported a significant detection with a hard
spectrum that connects smoothly with the TeV emission, while
\citet{Guo_2018} independently found a slightly extended GeV source
with a similarly hard spectrum. A later reanalysis by \citet{Cui2019}
revealed a more complex picture, in which the GeV emission is likely
composed of a hard component associated with the SNR and a softer,
compact component, denoted S0, toward the molecular-cloud core (MC-core) region.
Previous broadband studies have shown that both leptonic and hadronic
contributions can reproduce different aspects of the observed emission,
with the inferred particle populations depending on the adopted
environmental and spatial assumptions
\citep{HESS2011,Guo_2018,Cui2016,Cui2019}. Motivated by these
results and by the substantially increased LAT exposure, we perform
here a new analysis of approximately 16~yr of \textit{Fermi}-LAT observations,
including dedicated tests of the source morphology and a spatial
separation of the MC-core/S0 component from the remaining SNR emission.

The degeneracy between leptonic and hadronic scenarios outlined above is
unlikely to be resolved with current-generation instruments alone.
The Cherenkov Telescope Array Observatory
(CTAO, \citealt{2019scta.book.....C}) will provide improved
sensitivity, angular resolution, and energy coverage from tens
of GeV to beyond 100~TeV, enabling detailed morphological and
spectral studies of SNRs in this energy regime.

In this work, we model the broadband spectral energy distribution (SED)
of HESS~J1731$-$347 to constrain the underlying particle distributions and to assess the
relative contributions of leptonic and hadronic emission processes. We
combine archival radio, X-ray, and H.E.S.S. measurements with the new
\textit{Fermi}-LAT spectra and constrain the model parameters with Markov
Chain Monte Carlo (MCMC) sampling within the \texttt{naima} framework
\citep{Zabalza2015}. We then use dedicated three-dimensional simulations
of CTAO observations, performed with the \texttt{Gammapy} package
\citep{2023A&A...678A.157D}, to assess how future CTAO data can constrain
the TeV spectral shape and test the predictions of the specific best-fit
models considered here.
The paper is organized as follows. Section~\ref{sec:data} describes the
multi-wavelength dataset and the \textit{Fermi}-LAT analysis, and
Section~\ref{sec:fermi_results} presents the LAT results and defines the two
$\gamma$-ray spatial configurations used in the modeling.
Section~\ref{sec:model} describes the broadband modeling framework, and
Section~\ref{sec:scenarios} presents the leptonic and lepto-hadronic results
for both spatial configurations. Section~\ref{sec:ctao} presents the CTAO
simulations, Section~\ref{sec:discussion} discusses the physical implications
and limitations of our results, and Section~\ref{sec:conclusions} summarizes
our conclusions.

% ============================================================
\section{Multi-wavelength Data and \textit{Fermi}-LAT Analysis}
\label{sec:data}
% ============================================================

\begin{figure}
    \includegraphics[width=\hsize,clip]{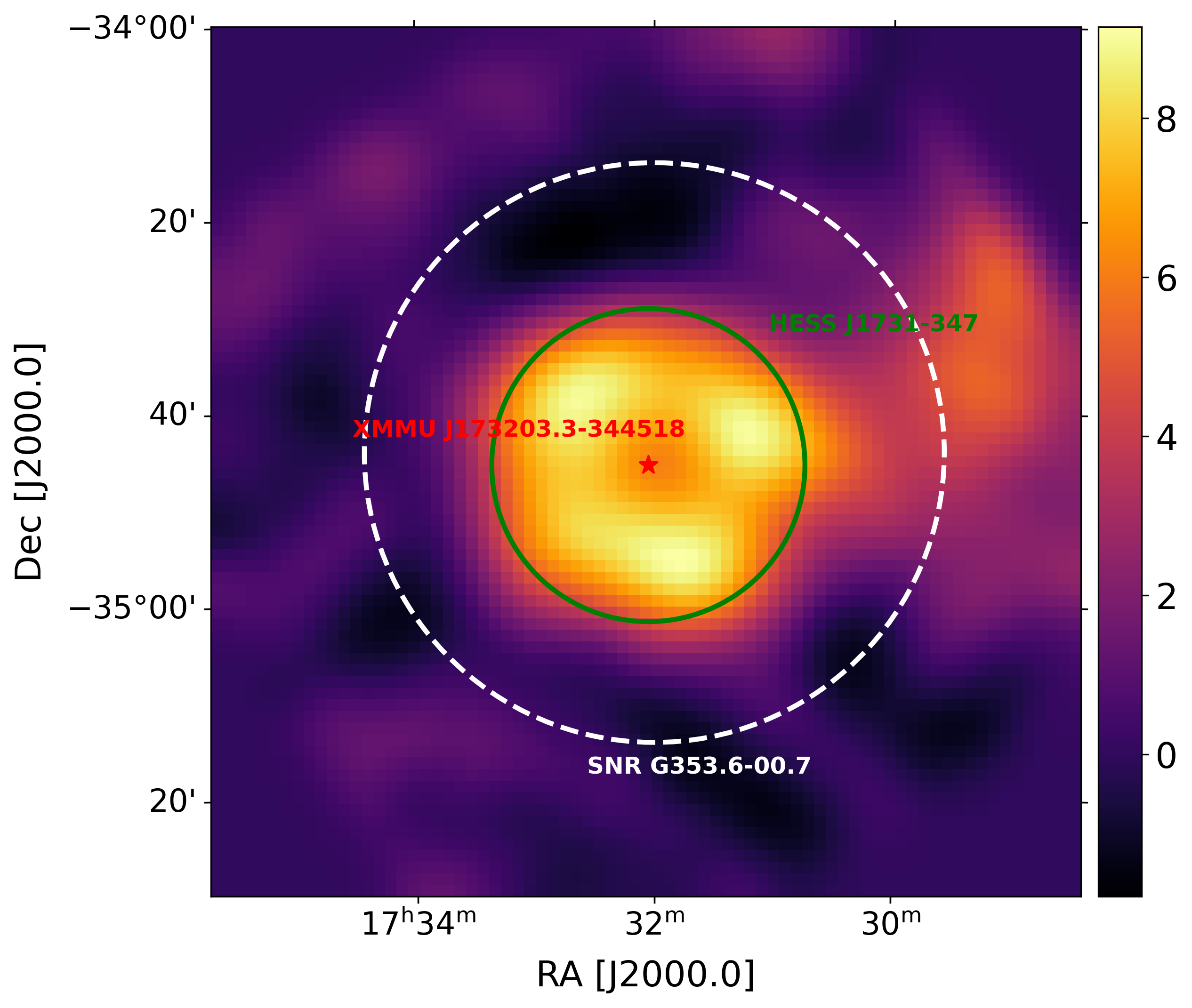}
    \caption{H.E.S.S. VHE $\gamma$-ray significance map of the
HESS~J1731$-$347 region \citep{2018A&A...612A...1H},
smoothed with a Gaussian kernel of width $0.04^{\circ}$.
The color scale gives the significance in units of $\sigma$.
The red star marks the position of the central compact object
XMMU~J173203.3$-$344518 \citep{Halpern2010,Tian2010}.
The white circle indicates the radio extent of SNR G353.6$-$0.7,
with a radius of $30'$ \citep{Tian2008}, while the green circle
shows the $1\sigma$ extension of the TeV source measured by
H.E.S.S. \citep{HESS2011,2018A&A...612A...1H}.
}
    \label{fig:signif_map}
\end{figure}

To constrain the physical parameters of the relativistic particle
populations accelerated in HESS~J1731$-$347, we constructed a broadband
SED combining archival radio, X-ray, and TeV measurements with a new
analysis of \textit{Fermi}-LAT observations presented in this work. The resulting
dataset spans approximately nineteen orders of magnitude in photon energy,
from radio frequencies to VHE $\gamma$-rays. The different spectral bands
constrain complementary portions of the underlying particle distributions
and radiative processes, allowing the leptonic and lepto-hadronic scenarios
to be tested within a common broadband framework.
Figure~\ref{fig:signif_map} provides an overview of the HESS~J1731$-$347
region in VHE $\gamma$-rays and shows the main spatial
features relevant to the multi-wavelength analysis.

\subsection{Radio, X-ray, and TeV Data}
\label{sec:radio_rX_data}

In the radio domain, we included the flux densities at
325 and 610~MHz obtained with the Giant Metrewave Radio Telescope
(GMRT) and reported by \citet{Nayana2017}, together with the
1420~MHz measurement from the VLA Galactic Plane Survey (VGPS)
and Australia Telescope Compact Array (ATCA)/Parkes observations
\citep{Tian2008}. These measurements sample the optically thin
synchrotron emission from the shell and constrain the low-energy
slope and normalization of the electron distribution in combination
with the magnetic field.

The X-ray emission of HESS~J1731$-$347 is dominated by a featureless
non-thermal continuum across the entire shell, consistent with synchrotron
radiation from relativistic electrons accelerated at the forward shock.
We adopted the spectrum derived from the deep XMM-Newton observations
analyzed by \citet{Doroshenko2017}, which probes the high-energy portion
of the electron distribution and provides an important constraint on its
spectral cutoff.

For the TeV component of the broadband SED, we adopted the differential
$\gamma$-ray flux measurements of HESS~J1731$-$347 reported by
\citet{HESS2011}. These measurements constrain the high-energy behavior
of the parent particle distributions, including the location of their
spectral cutoffs. The way these points are combined with the spatially
separated LAT spectra is described in Section~\ref{sec:configurations}.

\subsection{\textit{Fermi}-LAT Data Analysis}
\label{sec:fermi_data}

We performed a new analysis of approximately 16~yr of
\textit{Fermi}-LAT observations toward HESS~J1731$-$347, covering the interval from 2008 August 4 to
2024 August 2. The analysis was carried out with \texttt{Fermipy} v1.4.3
\citep{Wood2017} and Fermitools v2.5.1. \texttt{Fermipy} was used to construct
and optimize the region-of-interest (ROI) model, perform the
binned-likelihood analysis, test alternative spatial descriptions of
the source, inspect the residual emission, and derive the final
SEDs.

We selected Pass~8 SOURCE-class events in the energy range
1 to 300~GeV, adopting \texttt{evclass=128}, \texttt{evtype=3}, a
maximum zenith angle of $90^\circ$, and the
P8R3\_SOURCE\_V3 instrument response functions. Energy-dispersion
corrections were enabled throughout the analysis. The ROI was centered
at
$(\mathrm{R.A.},\mathrm{Decl.}) =
(263.0125^\circ,-34.7555^\circ)$
and had an angular size of $10^\circ$. A spatial bin size of
$0.1^\circ$ and eight energy bins per decade were adopted. The
Galactic diffuse emission and isotropic background were described by
\texttt{gll\_iem\_v07.fits} and
\texttt{iso\_P8R3\_SOURCE\_V3\_v1.txt}, respectively. The initial
source model was constructed from the 16~yr \textit{Fermi}-LAT source
list (FL16Y; \citealt{fermi_16,Abdollahi2022}), which contains FL16Y J1732.0$-$3445 at an angular
separation of only $\simeq0.01^\circ$ from the reference center of
HESS~J1731$-$347.

To ensure a consistent comparison among the different spatial hypotheses,
the same treatment of background and neighboring sources was adopted in all
likelihood fits. The normalizations of sources within $3^\circ$ of
HESS~J1731$-$347 with a test statistic (TS) $\geq 9$ were allowed to vary,
while the spectral parameters of the target and the normalizations of the
Galactic and isotropic diffuse components were left free. The likelihood
optimization was performed with the \texttt{Fermipy} \texttt{fit} routine.
Throughout the analysis, the TS is defined as
\begin{equation}
{\rm TS} = 2 \ln \left(\frac{\mathcal{L}_1}{\mathcal{L}_0}\right),
\end{equation}
where $\mathcal{L}_1$ and $\mathcal{L}_0$ denote the maximum likelihoods
obtained with and without the source component under consideration,
respectively.

We first considered a point-source description of the GeV emission
and then tested extended spatial models with the \texttt{Fermipy}
\texttt{extension} routine. The extended hypotheses included a
uniform disk, an analytical shell based on the morphology reported in the
H.E.S.S. Galactic Plane Survey (HGPS), and an empirical template constructed
from the H.E.S.S. excess map. This comparison was used to assess the
sensitivity of the LAT likelihood to the assumed morphology and to define a
spatial model suitable for a consistent comparison between the GeV and TeV
measurements.

The empirical H.E.S.S.-based template was constructed from the publicly available smoothed excess map of HESS~J1731$-$347 \citep{HESS2011,2018A&A...612A...1H}. A local background level was estimated from an annulus extending from $0.45^\circ$ to $0.75^\circ$ from the source center and subtracted from the map, after which negative residual values were set to zero. To restrict the spatial model to the SNR region while avoiding an abrupt boundary, we applied a radial raised-cosine window defined as
\[
w(r)=
\begin{cases}
1, & r \leq 0.35^\circ,\\[4pt]
\frac{1}{2}\left[1+\cos\left(
\pi\frac{r-0.35^\circ}{0.10^\circ}
\right)\right],
& 0.35^\circ < r < 0.45^\circ,\\[4pt]
0, & r \geq 0.45^\circ,
\end{cases}
\]
where $r$ is the angular separation from the adopted SNR center. This window preserves the original map intensity within $0.35^\circ$ and smoothly suppresses the emission to zero between $0.35^\circ$ and $0.45^\circ$, preventing a sharp artificial edge in the spatial template. The resulting map was reprojected onto a Galactic TAN grid with $0.005^\circ$ pixels and normalized to unit integral. We refer to this product as an empirical H.E.S.S.-based spatial template, since it retains the smoothing and analysis imprint of the published excess map and should not be interpreted as a reconstruction of the intrinsic TeV surface-brightness distribution.

\begin{figure*}
    \centering

    \includegraphics[width=0.95\textwidth]
    {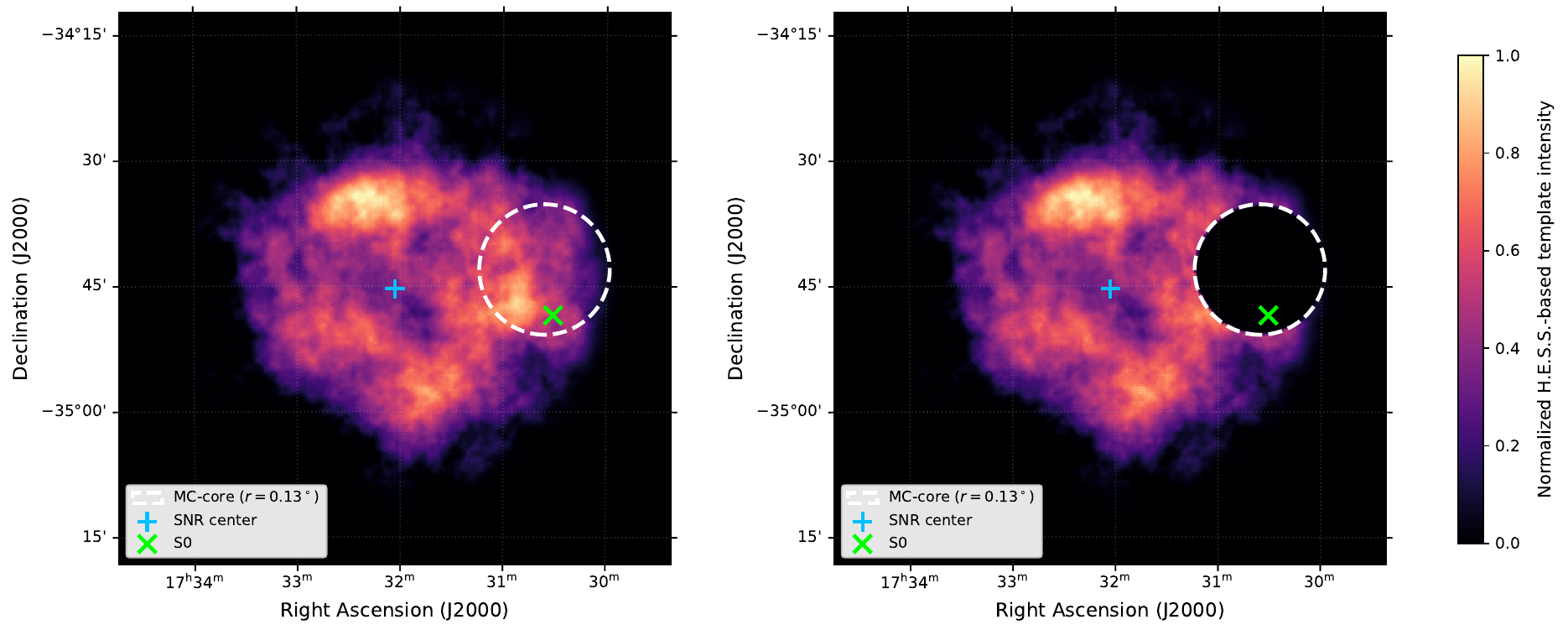}

    \vspace{-0.2cm}

    \begin{minipage}{0.46\textwidth}
        \centering
        \textbf{(a)} Full H.E.S.S.-based template
    \end{minipage}
    \hfill
    \begin{minipage}{0.46\textwidth}
        \centering
        \textbf{(b)} H.E.S.S.-based template excluding MC-core
    \end{minipage}

    \caption{H.E.S.S.-based spatial templates adopted in the
    \textit{Fermi}-LAT analysis of HESS~J1731$-$347. Panel (a) shows the full
    empirical template constructed from the smoothed H.E.S.S. excess map,
    whereas panel (b) shows the corresponding template after excision of the
    MC-core region. The dashed circle indicates the adopted MC-core
    region, with radius $r=0.13^\circ$. The cyan plus marks the reference
    center of HESS~J1731$-$347, and the green cross indicates the
    position of the GeV component S0 adopted from \citet{Cui2019}.
    Both templates were normalized to unit integral before being used in
    the LAT likelihood analysis.}
\label{fig:hess_templates}

\end{figure*}

To investigate how the S0 component \citep{Condon_2017,Cui2019} affects
the spectrum attributed to HESS~J1731$-$347, we considered two spatial
configurations. In the first, the full H.E.S.S.-based
template was used to characterize the integrated GeV emission from the
SNR region, including the MC-core. In the second, the MC-core region was
excised from the template and S0 was modeled independently as a point
source. Following the MC-core definition adopted by \citet{Cui2019}, a
circular region of radius $0.13^\circ$ centered at
$(\mathrm{R.A.},\mathrm{Decl.}) =
(262.650^\circ,-34.7167^\circ)$
was removed from the H.E.S.S.-based template. The position adopted
for S0,
$(\mathrm{R.A.},\mathrm{Decl.}) =
(262.629^\circ,-34.808^\circ)$,
was also taken from \citet{Cui2019}. After the excision, the remaining
SNR template was renormalized to unit integral before being used in
the likelihood analysis. The full and MC-core-excised spatial
templates are shown in Figure~\ref{fig:hess_templates}.

For both spatial configurations, the LAT SED was extracted between
1 and 300~GeV with the \texttt{Fermipy} \texttt{sed} routine. The energy range
was divided into five logarithmically spaced intervals, with the
background parameters refitted independently in each bin and a local
spectral index adopted for the target. Flux measurements were retained
for bins with ${\rm TS}\geq4$, whereas 95\% confidence-level upper limits
were derived for bins with ${\rm TS}<4$.

% ============================================================
\section{\textit{Fermi}-LAT Results and $\gamma$-ray Spatial Configurations}
\label{sec:fermi_results}
% ============================================================

In this section we characterize the morphology and spectrum of the GeV
emission toward HESS~J1731$-$347, quantify the contribution of the
MC-core/S0 region, and define the two GeV-plus-TeV datasets used in the
broadband modeling.

\subsection{Morphology of the GeV Emission}
\label{sec:fermi_morphology}

Under the point-source hypothesis, the fit yielded
$\mathrm{TS}=38.16$, a photon index of
\textcolor{blue}{$\Gamma=2.25\pm0.17$}, and an integrated photon flux of
\textcolor{blue}{$F_{1-300\,{\rm GeV}} = (8.68 \pm 2.16)\times10^{-10}$
ph~cm$^{-2}$~s$^{-1}$}. Allowing the source position to vary resulted
in a displacement of only $0.0215^\circ$ relative to the adopted SNR
center. Because this offset is small compared with the angular scales
relevant to the LAT analysis, the nominal SNR position was retained
for the subsequent morphological tests.

The extension analysis with a uniform-disk morphology yielded
$\mathrm{TS}_{\rm ext}=14.11$, with a fitted 68\% containment radius of
$R_{68}=0.199^\circ$ and a 95\% upper limit of
$R_{68}<0.257^\circ$. We therefore regard the LAT data as
suggestive of spatial extension, but not as an unambiguous extension
measurement.

We then compared the point-source model with three fixed extended
spatial descriptions, namely the analytical HGPS shell, the uniform disk, and the
empirical H.E.S.S.-based template described in
Section~\ref{sec:fermi_data}. The source TS increases from 38.16 for the
point source to 51.22, 56.61, and 58.62 for the HGPS shell, uniform disk,
and H.E.S.S.-based template, respectively. The results of this comparison
are summarized in Table~\ref{tab:fermi_morphology}.

\begin{table}[t]
\centering
\caption{Comparison of the spatial models tested for the
\textit{Fermi}-LAT emission associated with HESS~J1731$-$347.
$\mathrm{RMS}_{\rm res}$ denotes the root-mean-square of the
residual-significance distribution evaluated within $0.5^\circ$
of the adopted SNR center.}
\label{tab:fermi_morphology}

\setlength{\tabcolsep}{3.5pt}
\renewcommand{\arraystretch}{1.05}

\begin{tabular}{lcccc}
\hline
\hline
Spatial model &
$\ln\mathcal{L}$ &
TS &
$\Gamma$ &
$\mathrm{RMS}_{\rm res}$ \\
\hline
Point source      & 1914490.282 & 38.16 & $2.25 \pm 0.17$ & 1.269 \\
HGPS shell        & 1914499.323 & 51.22 & $2.06 \pm 0.14$ & 1.022 \\
Uniform disk      & 1914502.226 & 56.61 & $2.02 \pm 0.13$ & 1.076 \\
H.E.S.S. template & 1914502.611 & 58.62 & $2.05 \pm 0.13$ & 0.936 \\
\hline
\end{tabular}
\end{table}

The uniform-disk and H.E.S.S.-based template models yield very
similar maximum likelihoods, with
$2\Delta\ln\mathcal{L}\simeq0.77$. Because these two hypotheses are
fixed, non-nested morphologies, we do not interpret this difference as a
formal significance, and the LAT data alone do not distinguish between
them. We adopt the H.E.S.S.-based template for the subsequent analysis
because it provides a common spatial definition for the GeV and TeV
emission and gives the lowest residual RMS among the morphologies tested.

\subsection{Separation of the SNR and MC-core/S0 Emission}
\label{sec:fermi_s0}

As an initial test, S0 was introduced as a point source while the
full H.E.S.S.-based template was retained for HESS~J1731$-$347. In
this configuration, the addition of S0 yielded only
$\mathrm{TS}_{\mathrm{add,S0}}=6.08$. Because the full template
contains non-zero intensity within the MC-core region, part of the
emission spatially coincident with S0 can be absorbed by the extended
SNR component, and this configuration does not provide a clean
spatial separation between the two components.

We then repeated the likelihood analysis with the
MC-core-excised template defined in
Section~\ref{sec:fermi_data}. With the SNR component alone, the
fit yielded $\ln\mathcal{L}=1914499.686$. After S0 was introduced as
an independent point source and the spectral parameters of both
components were refitted, the likelihood increased to
$\ln\mathcal{L}=1914505.988$, corresponding to
$\mathrm{TS}_{\mathrm{add,S0}}=12.60$. The larger contribution
attributed to S0 in this configuration is consistent with a reduced
spatial degeneracy between S0 and the extended SNR component
once the MC-core region is excised.

In the resulting two-component fit, the MC-core-excised SNR
component has $\mathrm{TS}=36.23$, a photon index of
$\Gamma_{\rm SNR}=1.97\pm0.16$, and an integrated photon flux of
$F_{\rm SNR}(1\mbox{ to }300~{\rm GeV})
=(1.01 \pm 0.25)\times10^{-9}$ ph~cm$^{-2}$~s$^{-1}$. The S0 component has
$\mathrm{TS}=14.97$, with $\Gamma_{\rm S0}=2.28\pm0.19$ and
$F_{\rm S0}(1\mbox{ to }300~{\rm GeV})
=(5.83 \pm 2.07)\times10^{-10}$ ph~cm$^{-2}$~s$^{-1}$.
The best-fit spectrum of S0 is nominally softer than that of the
SNR component, in qualitative agreement with \citet{Cui2019}. However, the
difference between the two photon indices corresponds to only
about $1.2\sigma$ when their statistical uncertainties are
combined, so we regard it as a trend rather than as independent evidence
for two distinct particle populations.

The residual-significance distribution of the final SNR+S0 model,
evaluated within $0.5^\circ$ of the SNR center, has an
RMS of 0.87, a mean of 0.031, and a maximum absolute residual
significance of $1.97\sigma$. At the SNR center and near the adopted
position of S0, the residual significances are approximately
$-0.22\sigma$ and $+0.24\sigma$, respectively, so no prominent localized
residual remains at either position once both components are included.
These quantities are used as diagnostics of the spatial model and not as a
formal global goodness-of-fit statistic.

\subsection{GeV Spectra and the Adopted Spatial Configurations}
\label{sec:configurations}

\begin{figure}[htbp]
    \centering
    \includegraphics[width=\columnwidth]
    {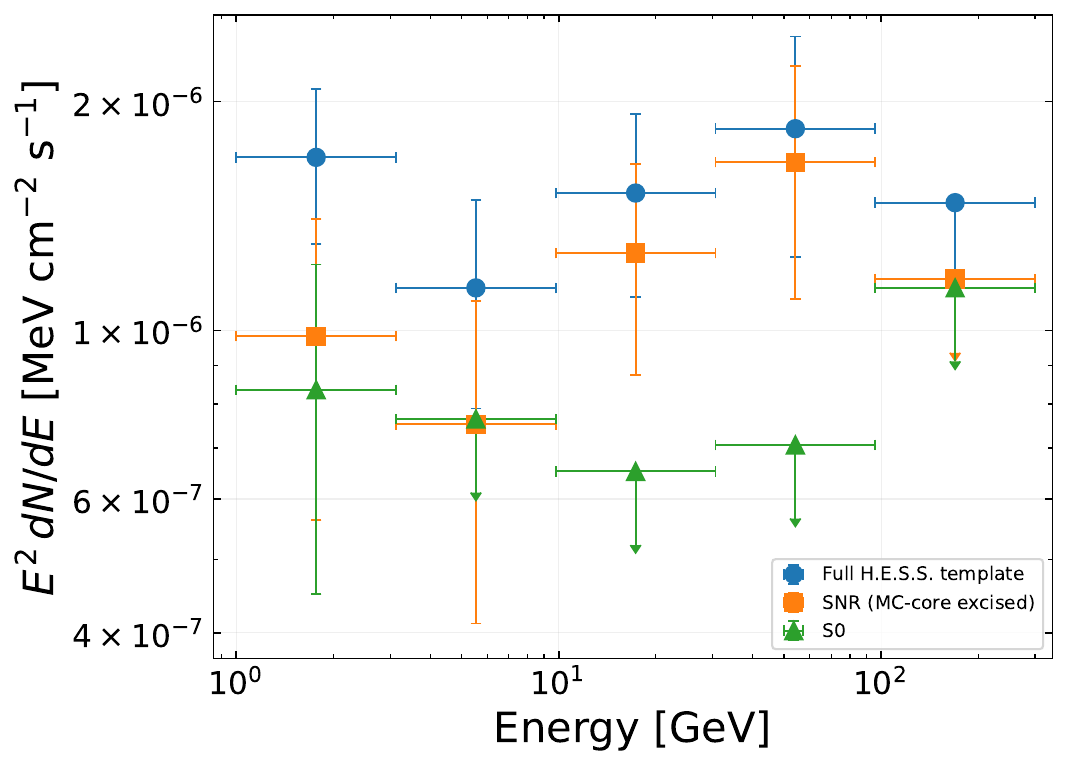}
    \caption{\textit{Fermi}-LAT spectral energy distributions of the
    HESS~J1731$-$347 region derived from the 16~yr analysis presented
    in this work. The spectrum obtained with the full H.E.S.S.-based
    template is compared with that of the SNR after excision of the
    MC-core region and with the independently fitted S0 component.
    Horizontal error bars indicate the boundaries of the energy bins.
    Flux measurements are shown for bins with $\mathrm{TS}\geq4$, whereas
    downward arrows denote 95\% confidence-level upper limits.}
    \label{fig:fermi_seds}
\end{figure}

Figure~\ref{fig:fermi_seds} compares the LAT spectrum obtained with the
full H.E.S.S.-based template with those of the MC-core-excised SNR
and of the independently fitted S0 component. The 1 to 300~GeV range was
divided into five logarithmically spaced intervals, with boundaries at
1.000, 3.129, 9.792, 30.639, 95.874, and 300~GeV. Both the integrated
and the MC-core-excised SNR spectra yield flux measurements in the first
four bins and a 95\% confidence-level upper limit in the highest-energy
interval. In contrast, S0 satisfies the $\mathrm{TS}\geq4$ criterion only
in the lowest-energy bin, and the remaining four bins provide upper limits.

The treatment of the MC-core region matters most at the
low-energy end of the LAT spectrum. In the four bins where
both SNR configurations yield flux measurements, the spectrum derived
with the full template exceeds that obtained after
MC-core excision by approximately 72\%, 51\%, 20\%, and 11\%,
in order of increasing energy. This progressively smaller difference is
consistent with a relatively larger contribution from the spectrally
softer S0 component at lower GeV energies. It should not be read as a
direct spectral decomposition of the MC-core emission, however, because the
two SNR spectra are obtained under different spatial hypotheses and the
LAT point-spread function is broader at the lower end of the energy range.

To keep the GeV and TeV data consistent with each spatial configuration,
we built two corresponding representations of the H.E.S.S. spectrum. For the
integrated configuration, the published H.E.S.S. points were used
without modification. For the MC-core-excised configuration, the same
points were multiplied by a factor of 0.8423, which is the fraction of the
integrated intensity retained in the processed H.E.S.S.-based template
after excision of the MC-core region. This rescaling approximates the TeV
emission from the same region used to derive the MC-core-excised LAT
spectrum and is not an independent measurement of the H.E.S.S. spectrum
outside the MC-core. It assumes that the fractional contribution of the
excised region is approximately independent of energy over the H.E.S.S.
range, so any energy-dependent variation of the TeV morphology is not
captured. For comparison, \citet{Cui2019} estimated that approximately 88\%
of the H.E.S.S. emission lies outside the MC-core region. The difference
relative to the 84.23\% retained here arises from the specific
H.E.S.S. map product, template processing, and intensity-integration
procedure adopted in this work, and not from a different definition of the
excised region.

The MC-core-excised LAT and H.E.S.S. spectra together form the fiducial
$\gamma$-ray dataset for the broadband modeling, because they reduce the
contribution of the spatially distinct S0/MC-core region and are more
directly associated with the remaining SNR emission. The integrated
full-template dataset is retained as a consistency check on the effect of
the spatial decomposition. In both cases, the radio and X-ray measurements
of Section~\ref{sec:radio_rX_data} are unchanged.

% ============================================================
\section{Broadband Modeling Framework}
\label{sec:model}
% ============================================================

The broadband emission of HESS~J1731$-$347 is modeled by linking
the observed SED to the underlying
relativistic particle populations. The radiative processes considered
in this work are synchrotron emission, IC scattering
on multiple target photon fields, and neutral-pion ($\pi^{0}$) decay
from inelastic proton-proton interactions, all computed with the
\texttt{naima} package \citep{Zabalza2015}. This framework allows the
purely leptonic and lepto-hadronic interpretations to be evaluated
consistently using the same multi-wavelength dataset.

The source distance is fixed at $d = 3.2$~kpc in both
emission scenarios. For the lepto-hadronic model, the
ambient hydrogen density is fixed at
$n_{\mathrm{H}} = 20$~cm$^{-3}$, an assumed
target density for proton-proton interactions in the
spatially inhomogeneous environment of HESS~J1731$-$347
\citep{Fukuda2014}. This density is an environmental assumption and is
not constrained by the broadband fit. The magnetic field
and the parameters of the primary particle
distributions are inferred through the fitting procedure,
whereas the total electron and proton energy contents
are derived from the fitted distributions and are not
sampled as independent free parameters.

\subsection{Particle Distributions and Emission Scenarios}
\label{sec:particle_distribution}

Before adopting the final particle parametrization, we also
tested a log-parabolic distribution. This form did not describe the radio
emission and the $\gamma$-ray spectral peak simultaneously and
systematically underpredicted the low-frequency GMRT
measurements. We therefore adopt an exponentially cutoff power-law
(ECPL) distribution throughout the analysis.
The ECPL form is physically grounded
in first-order Fermi acceleration at astrophysical
shock fronts, which naturally yields an approximately power-law
particle spectrum~\citep{Bell1978,Blandford1987}. At high energies,
departures from a pure power law can arise from the interplay between
the finite acceleration timescale, radiative cooling, particle escape,
and shock-geometry effects, including the role of obliquity in
regulating the maximum attainable energy~\citep{2026ApJ..1002...97P}.
This high-energy suppression is therefore modeled phenomenologically
by an exponential rollover~\citep{Reynolds1998,Zirakashvili2007}.
For each particle species $i=e,p$, the differential energy
distribution is written as
\begin{equation}
\frac{dN_i}{dE}
=
A_i
\left(
\frac{E}{E_0}
\right)^{-\alpha_i}
\exp
\left[
-\left(
\frac{E}{E_{{\rm cut},i}}
\right)^{\beta_i}
\right],
\label{eq:ecpl}
\end{equation}
where $A_i$ is the normalization, $\alpha_i$ is the spectral index,
and $E_{{\rm cut},i}$ is the cutoff energy. The normalization is
defined at the common reference energy $E_0=1$~TeV, and we fix
$\beta_i=1$ for both electrons and protons, corresponding to a
simple exponential cutoff. For strong non-relativistic shocks, standard
DSA predicts indices close to $\alpha\simeq2$ under idealized conditions
\citep{Bell1978,Blandford1987}. Here, however, the spectral indices and
cutoff energies are inferred directly from the broadband data within the
prior ranges listed in Table~\ref{tab:priors}. The total non-thermal energy
contents $W_e$ and $W_p$ of the electron and proton populations are
computed by integrating Equation~(\ref{eq:ecpl}) above a minimum particle
energy of $E_{\rm min}=1$~GeV.

In the purely leptonic model, the radio and X-ray emission is produced
by synchrotron radiation from relativistic electrons, and the $\gamma$-ray
emission is attributed entirely to IC scattering by the same electron
population. The IC emission includes the cosmic microwave
background (CMB) together with far-infrared (FIR) and near-infrared
(NIR) components of the interstellar radiation field (ISRF), computed with the \texttt{GALPROP} code. At the Galactic location of
HESS~J1731$-$347, we adopt an FIR component with
$T=40$~K and an energy density of
$u_{\rm FIR}=0.313~{\rm eV\,cm^{-3}}$, and an NIR component with
$T=3000$~K and
$u_{\rm NIR}=0.663~{\rm eV\,cm^{-3}}$ \citep{Porter2008,Porter2022}. These photon fields are
kept fixed in all model configurations.

In the lepto-hadronic model, the synchrotron and IC components are
supplemented by $\gamma$-ray emission from neutral-pion decay
following inelastic proton-proton interactions. Radiation from secondary
electrons and positrons produced in charged-pion decays is not included
explicitly in the present one-zone calculation, and the implications of this
limitation are discussed in Section~\ref{sec:discussion}. The relative
normalization of the two populations is parametrized through the
electron-to-proton normalization ratio
\begin{equation}
K_{\rm ep} = \frac{A_e}{A_p},
\label{eq:kep}
\end{equation}
where $A_e$ and $A_p$ are the normalizations of the electron and
proton ECPL distributions at $E_0=1$~TeV. $K_{\rm ep}$ is therefore a ratio
at the pivot energy and should not be interpreted as the ratio of the total
energy contents, $W_e/W_p$. In the MCMC analysis, $K_{\rm ep}$ is sampled
as a free parameter, and the proton normalization is obtained from
$A_p=A_e/K_{\rm ep}$.

\subsection{Bayesian Inference and Model Comparison}
\label{sec:mcmc}

The posterior distributions of the model parameters were explored
within a Bayesian framework \citep{MacKay2003,Gelman2013} using
MCMC sampling with the \texttt{emcee} affine-invariant ensemble sampler
\citep{Goodman2010,Foreman-Mackey2013}, as implemented in
\texttt{naima} \citep{Zabalza2015}. For a model $M$ with parameter
vector $\boldsymbol{\theta}$ and dataset $D$, the posterior
distribution is
\begin{equation}
P(\boldsymbol{\theta}\mid D,M)
\propto
\mathcal{L}(D\mid\boldsymbol{\theta},M)\,
\pi(\boldsymbol{\theta}),
\end{equation}
where $\mathcal{L}$ is the likelihood and
$\pi(\boldsymbol{\theta})$ is the prior distribution. For flux
measurements, the likelihood is evaluated using the reported uncertainties.
Spectral bins without a significant detection, in particular the
\textit{Fermi}-LAT bins with $\mathrm{TS}<4$, are supplied to
\texttt{naima} as 95\% confidence-level upper limits and not as ordinary
flux measurements.

Uniform priors over finite intervals were adopted for the sampled
parameters (Table~\ref{tab:priors}). The electron normalization and cutoff
energy, and the proton cutoff energy in the lepto-hadronic model, are
sampled through their base-10 logarithms, so the quoted priors are uniform in
these logarithmic variables. The prior on $K_{\rm ep}$ is uniform in linear
space. An initial burn-in phase was discarded before constructing the
marginalized posterior distributions. Parameter estimates are
reported as the median of the marginalized posteriors, with
uncertainties given by the 16th and 84th percentiles, which correspond to
68\% credible intervals. The maximum-likelihood solution is used separately
to evaluate the information criteria described below. Fractional residuals,
defined as $(F_{\rm data}-F_{\rm model})/F_{\rm model}$, are shown in the
broadband SED figures as an additional visual diagnostic of the fit
quality.

\begin{deluxetable*}{lcc}
\tablecaption{Prior ranges adopted in the broadband MCMC modeling of
HESS~J1731$-$347.\label{tab:priors}}
\tablehead{
\colhead{Parameter} &
\colhead{Leptonic} &
\colhead{Lepto-hadronic}
}
\startdata
$\log_{10}(A_e/{\rm eV}^{-1})$
    & $[30,\,38]$
    & $[30,\,35]$ \\
$\alpha_e$
    & $[1.5,\,3.0]$
    & $[1.5,\,2.5]$ \\
$\log_{10}(E_{\rm cut,e}/{\rm TeV})$
    & $[0.5,\,3.0]$
    & $[0.5,\,3.0]$ \\
$B~(\mu{\rm G})$
    & $[1,\,100]$
    & $[10,\,150]$ \\
$K_{\rm ep}$
    & \nodata
    & $[10^{-4},\,10^{-1}]$ \\
$\alpha_p$
    & \nodata
    & $[1.5,\,2.5]$ \\
$\log_{10}(E_{\rm cut,p}/{\rm TeV})$
    & \nodata
    & $[0.5,\,3.0]$
\enddata

\tablecomments{
The normalization parameters are defined at the common reference
energy $E_0=1$~TeV. In the lepto-hadronic model,
$K_{\rm ep}=A_e/A_p$ is sampled uniformly in linear space, and the
proton normalization is derived as $A_p=A_e/K_{\rm ep}$ and is not
sampled independently. Entries marked with \nodata\ do not apply to
the purely leptonic model.
}
\end{deluxetable*}

Both radiative scenarios are fitted separately to the two
$\gamma$-ray spatial configurations defined in
Section~\ref{sec:configurations}. To compare the leptonic and lepto-hadronic
models within a given configuration, we use the Bayesian Information
Criterion (BIC, \citealt{Schwarz1978}),
\begin{equation}
{\rm BIC} = k\ln n - 2\ln\hat{\mathcal{L}},
\label{eq:bic}
\end{equation}
where $k$ is the number of free parameters, $n$ is the number of
data points included in the fit, and $\hat{\mathcal{L}}$ is the
maximum likelihood. The BIC balances the improvement in the maximized
likelihood against the complexity introduced by extra free parameters, and
lower values indicate the preferred description. Differences in BIC are
interpreted following the scale of \citet{Kass1995}, bearing in mind
that the BIC is an asymptotic approximation to the model evidence and not a
direct calculation of the Bayes factor. For each spatial configuration, we
define
\begin{equation}
\Delta{\rm BIC}
=
{\rm BIC}_{\rm lep}
-
{\rm BIC}_{\rm LH},
\label{eq:delta_bic}
\end{equation}
where ${\rm BIC}_{\rm lep}$ and ${\rm BIC}_{\rm LH}$ refer to the
purely leptonic and lepto-hadronic models, respectively, so positive
values indicate a lower BIC for the lepto-hadronic model.
Because the two spatial configurations use different $\gamma$-ray datasets,
their absolute likelihoods and BIC values are never compared with one
another. Model comparison is performed only between the two radiative
scenarios within the same configuration.

As a complementary goodness-of-fit diagnostic, we also compute a
chi-square statistic using detected flux measurements only,
\begin{equation}
\chi^2_{\rm det}
=
\sum_{j=1}^{N_{\rm det}}
\left[
\frac{F_j-M_j}{\sigma_j}
\right]^2,
\end{equation}
where $F_j$ and $M_j$ are the measured and model-predicted fluxes and
$N_{\rm det}$ is the number of detected flux measurements. Upper limits
remain in the likelihood analysis but are excluded from
$\chi^2_{\rm det}$. For asymmetric uncertainties, the lower uncertainty is
adopted when the model lies below the measured flux, and the upper
uncertainty when it lies above. The number of degrees of freedom is
$\nu=N_{\rm det}-k$, and the reduced statistic is
$\chi^2_{\nu,{\rm det}}=\chi^2_{\rm det}/\nu$. These quantities are
complementary diagnostics only, and the likelihood and BIC remain the basis
of the formal model comparison.

% ============================================================
\section{Broadband Modeling Results}
\label{sec:scenarios}
% ============================================================

We first present the fiducial MC-core-excised configuration and then
repeat the analysis with the integrated full-template dataset to assess how
the inferred particle properties depend on the spatial treatment. Within
each configuration, the purely leptonic and lepto-hadronic models are
fitted and compared independently.

\subsection{Fiducial MC-core-excised Configuration}
\label{model_noMC}

In the purely leptonic scenario, the MC-core-excised spectrum is
described by an electron population with $\alpha_e=2.35\pm0.02$, a cutoff 
energy of $E_{\rm cut,e}=29.3^{+2.2}_{-2.0}$ TeV, and a normalization of 
$A_e=(4.36\pm0.20)\times10^{34}$ eV$^{-1}$, with a magnetic field of 
$B=15.1\pm0.7$ $\mu$G. The corresponding total electron energy above 1 GeV is 
$W_e=(2.15^{+0.19}_{-0.18})\times10^{48}$ erg. A cutoff of several tens of TeV is 
required because the same electrons must reach high enough energies to reproduce 
the non-thermal X-ray emission while sustaining an IC component throughout the GeV 
and TeV range. For a fixed synchrotron flux, a lower magnetic field
requires a larger electron population, which in turn increases the IC
contribution. This coupling between $B$ and the electron normalization is
central to a one-zone leptonic interpretation, in which the same particles
must account for both the radio-to-X-ray and the $\gamma$-ray emission.

The inferred magnetic field is lower, and the electron spectrum softer,
than in some previous broadband models. For example, \citet{Guo_2018}
obtained $B\simeq27$ to $29~\mu$G and $\alpha_e\simeq1.6$ to $1.8$ for their
purely leptonic models. These differences should be considered in the
context of the different datasets, spatial treatments, and modeling
assumptions adopted in the two analyses.

When a relativistic proton population is included, the
electron component changes substantially. The fit yields $
\alpha_e=2.08\pm0.03$, $E_{\rm cut,e}=11.3^{+1.8}_{-1.4}$ TeV, and $A_e=(1.3^{+0.7}
_{-0.4})\times10^{33}$ eV$^{-1}$, while the magnetic field increases to 
$B=84.9^{+27.5}_{-22.9}$ $\mu$G. The fitted electron-to-proton normalization 
ratio is $K_{\rm ep}=(3.5^{+2.0}_{-1.2})\times10^{-3}$, which corresponds to a proton
normalization of $A_p = (3.72 \pm 0.48) \times10^{35}$~eV$^{-1}$. The proton
population is considerably harder than the electron distribution, with
$\alpha_p=1.69 \pm 0.10$ and $E_{\rm cut,p} = 23.5^{+10.1}_{-5.9}$~TeV. The inferred total energies above 1 GeV 
are $W_e=(2.25^{+1.04}_{-0.65})\times10^{46}$ erg and $W_p=(4.37^{+0.64}_{-0.51})\times10^{48}$ erg.
\citet{Guo_2018} also found hard proton spectra in their hybrid modeling of
HESS~J1731$-$347, with $\alpha_p\simeq1.5$ to $1.7$ depending on the adopted
cutoff shape, together with magnetic fields of approximately
$80$ to $87~\mu$G, similar to the value obtained here.

\begin{figure*}[ht!]
\centering

\begin{minipage}{0.49\textwidth}
    \centering
    \includegraphics[width=\textwidth]{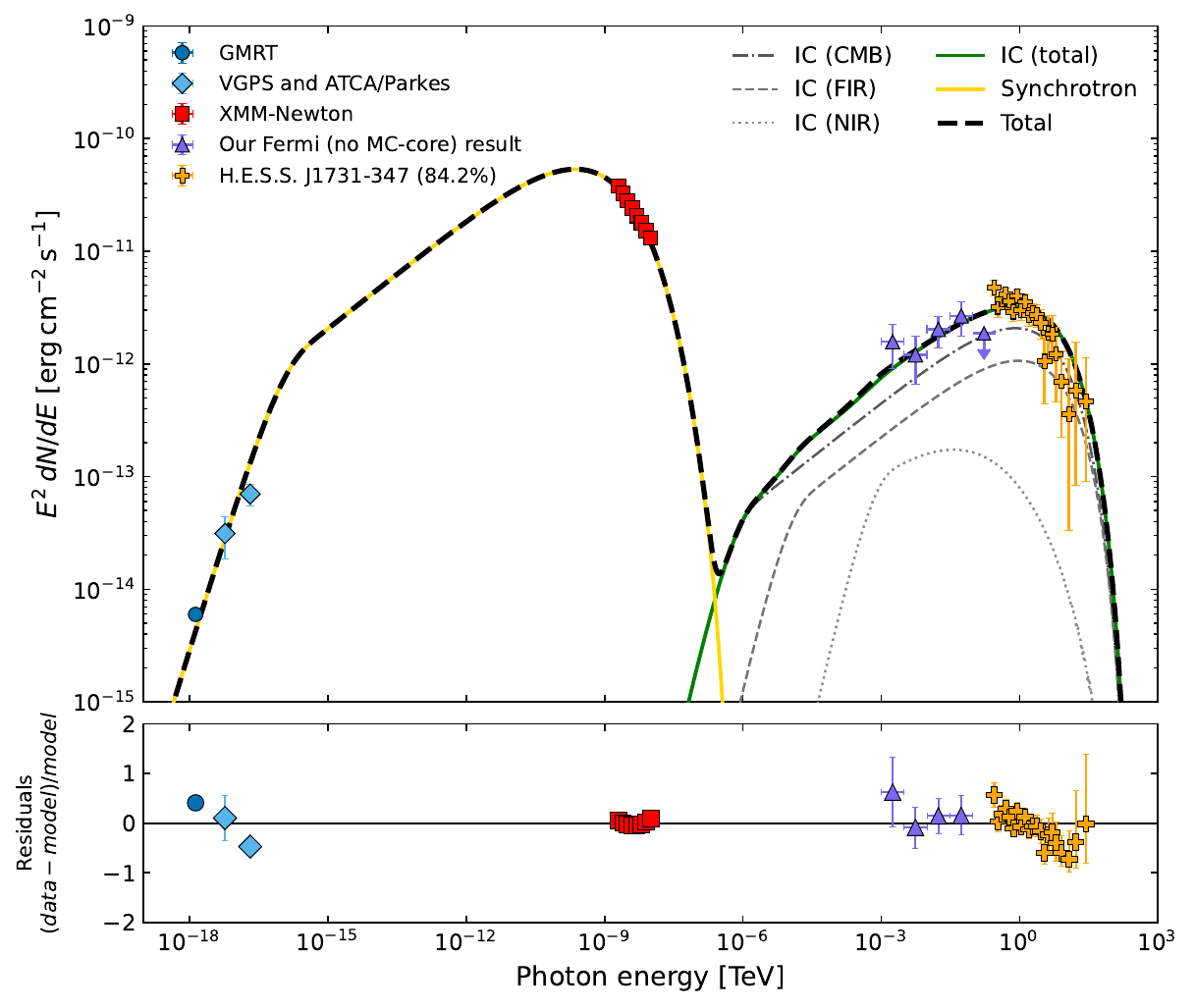}\\[0.3em]
    {\small (a) Leptonic Model}
\end{minipage}
\hfill
\begin{minipage}{0.49\textwidth}
    \centering
    \includegraphics[width=\textwidth]{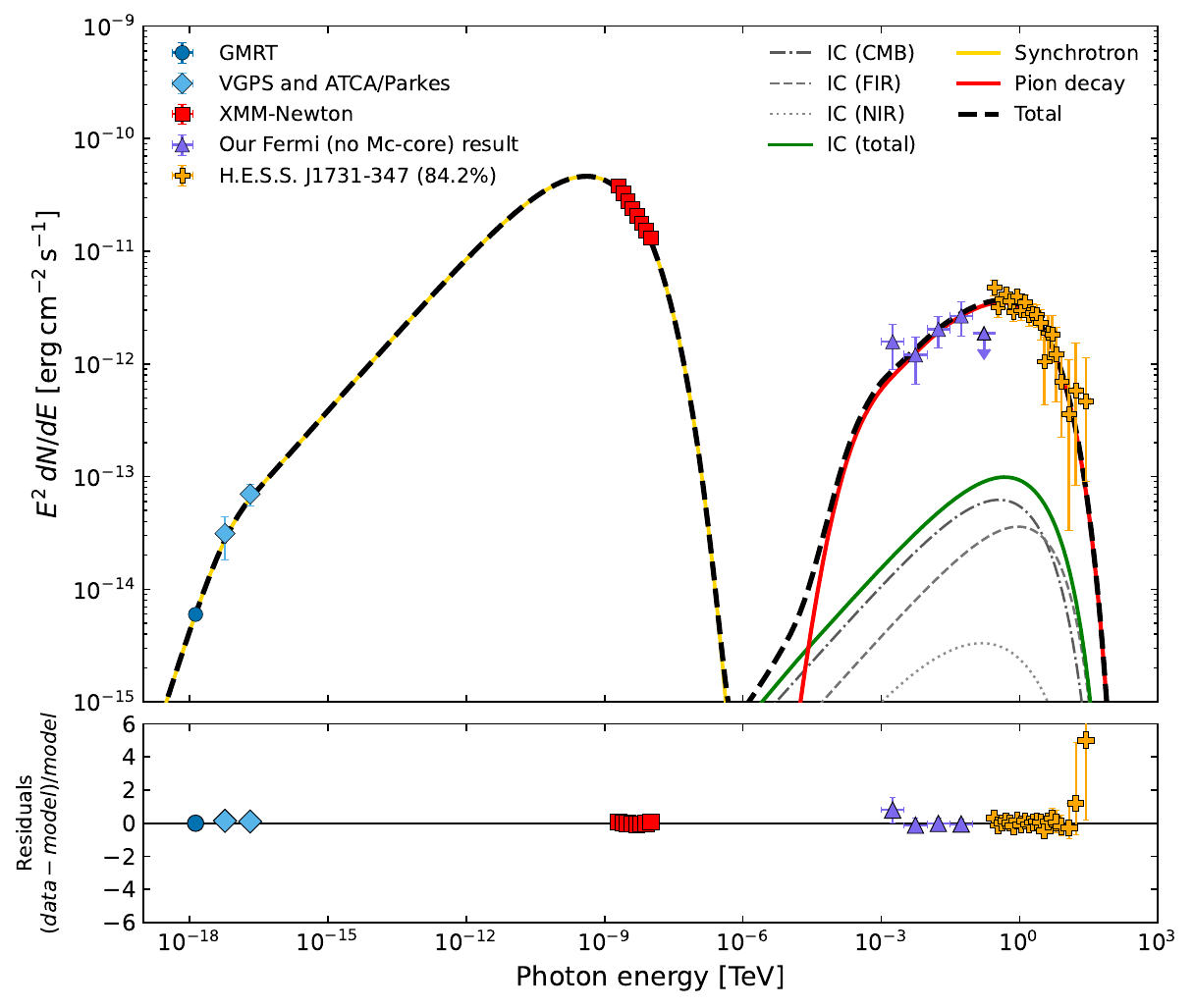}\\[0.3em]
    {\small (b) Lepto-hadronic Model}
\end{minipage}

\caption{Broadband spectral energy distributions of HESS~J1731$-$347 for the
fiducial MC-core-excised configuration. Panel (a) shows the purely
leptonic model, with synchrotron and IC emission, while
panel (b) shows the lepto-hadronic model including an additional
$\pi^0$-decay component from proton-proton interactions. The
individual IC contributions from the CMB, FIR, and NIR photon fields
are shown together with their sum. The lower panels show the fractional residuals, $(F_{\rm data}-F_{\rm model})/F_{\rm model}$.}
\label{fig:sed_no_MC}
\end{figure*}

Magnetic fields of several tens to $\sim100~\mu$G are commonly discussed
in the context of magnetic-field amplification at young SNR shocks
\citep{Reynolds2012,Vink2012}. In the present fit, the stronger field allows
the observed synchrotron emission to be produced with a smaller electron
normalization. This reduces the relative IC contribution and leaves a
larger fraction of the GeV and TeV emission to be supplied by $\pi^0$ decay.
The hard proton index deserves particular attention. Hard particle spectra
can arise under non-standard acceleration or environmental conditions, but
indices well below the canonical $\sim2$ of the simplest strong-shock DSA
picture are not trivial to reproduce, as also noted by \citet{Guo_2018}.
The hard proton spectrum should therefore be regarded as a requirement of
the present hadronic interpretation and not as independent evidence for
standard DSA. The broadband decompositions for both scenarios are shown in
Figure~\ref{fig:sed_no_MC}.

Because both models are fitted to the same dataset, their statistical
indicators can be compared directly. The leptonic model yields a maximum
log-likelihood of $\ln\hat{\mathcal{L}}=-33.85$ and
${\rm BIC}=82.14$, whereas the lepto-hadronic model reaches
$\ln\hat{\mathcal{L}}=-11.60$ with
${\rm BIC}=48.47$. The corresponding difference is
\begin{equation}
\Delta{\rm BIC}_{\rm noMC}
=
{\rm BIC}_{\rm lep}
-
{\rm BIC}_{\rm LH}
=
33.67.
\end{equation}
The improvement in maximum likelihood obtained with the lepto-hadronic
model therefore survives the penalty associated with its three additional
free parameters. On the scale of \citet{Kass1995}, a difference of this size
($\Delta{\rm BIC}>10$) corresponds to very strong evidence in favor of the
lepto-hadronic model within the MC-core-excised configuration. The
complementary diagnostic leads to the same ordering. The leptonic solution
gives $\chi^2_{\rm det}=63.09$ for 32 degrees of freedom
($\chi^2_{\nu,{\rm det}}=1.97$), whereas the lepto-hadronic solution
gives $\chi^2_{\rm det}=18.59$ for 29 degrees of freedom
($\chi^2_{\nu,{\rm det}}=0.64$).

The lower BIC of the lepto-hadronic model thus persists after the
S0/MC-core region is excluded. Within the adopted one-zone framework and
environmental assumptions, the statistical comparison favors a
non-negligible hadronic contribution to the broadband emission of the
remaining SNR. The inferred parameters and statistical indicators for both
scenarios are summarized in Table~\ref{tab:params_no_mc}.

The lower BIC of the lepto-hadronic model does not imply that it
reproduces every VHE measurement. Its prediction falls below the
highest-energy H.E.S.S. point, which shows the largest fractional residual
in Figure~\ref{fig:sed_no_MC}(b), although this point also carries the
largest uncertainty of the TeV spectrum. A steeper high-energy rollover of
the lepto-hadronic spectrum would therefore be in tension with this
measurement, and more precise data above $\sim20$~TeV are needed to test
it.

\begin{deluxetable*}{llcc}[ht!]
\tablecaption{Posterior parameter estimates and statistical indicators
for the MC-core-excised configuration of HESS~J1731$-$347,
considering the purely leptonic and lepto-hadronic emission models.
\label{tab:params_no_mc}}
\tablecolumns{4}
\tablehead{
\colhead{Parameter} & \colhead{Unit} &
\colhead{Leptonic} & \colhead{Lepto-hadronic}
}
\startdata
\noalign{\smallskip}
\multicolumn{4}{c}{Electron Distribution} \\
\noalign{\smallskip} \tableline \noalign{\smallskip}
Spectral index $\alpha_e$                           & \dots          & $2.35 \pm 0.02$               & $2.08 \pm 0.03$ \\
Cutoff energy $E_{\rm cut,e}$                       & TeV            & $29.3^{+2.2}_{-2.0}$           & $11.3^{+1.8}_{-1.4}$ \\
Normalization $A_e$                                 & eV$^{-1}$      & $(4.36 \pm 0.20)\times10^{34}$  & $(1.3^{+0.7}_{-0.4})
\times10^{33}$ \\
Total electron energy $W_e$ ($E > 1$~GeV)           & erg            & $(2.15^{+0.19}_{-0.18})\times10^{48}$          & $(2.25^{+1.04}_{-0.65})\times10^{46}$ \\
\noalign{\smallskip} \tableline \noalign{\smallskip}
\multicolumn{4}{c}{Proton Distribution} \\
\noalign{\smallskip} \tableline \noalign{\smallskip}
Spectral index $\alpha_p$                           & \dots          & \nodata                      & $1.69\pm 0.10$ \\
Cutoff energy $E_{\rm cut,p}$                       & TeV            & \nodata                      & $23.5\,^{+10.1}_{-5.9}$ \\
Normalization $A_p$                                 & eV$^{-1}$      & \nodata                      & $(3.72 \pm 0.48) \times 10^{35}$ \\
Total proton energy $W_p$ ($E > 1$~GeV)             & erg            & \nodata                      & $(4.37^{+0.64}_{-0.51})\times10^{48}$ \\
Electron-to-proton ratio $K_{\rm ep}$               & \dots          & \nodata                      & $(3.5^{+2.0}_{-1.2})\times10^{-3}$ \\
\noalign{\smallskip} \tableline \noalign{\smallskip}
\multicolumn{4}{c}{Environmental Conditions} \\
\noalign{\smallskip} \tableline \noalign{\smallskip}
Magnetic field $B$                                  & $\mu$G         & $15.1\pm 0.7$                & $84.9 \,^{+27.5}_{-22.9}$ \\
Distance $d$                                        & kpc            & $3.2$ (fixed)                & $3.2$ (fixed) \\
\noalign{\smallskip} \tableline \noalign{\smallskip}
\multicolumn{4}{c}{Statistical Indicators} \\
\noalign{\smallskip} \tableline \noalign{\smallskip}
Log-likelihood $\ln\mathcal{L}$                     & \dots          & $-33.85$                     & $-11.60$ \\
Bayesian Information Criterion, BIC                 & \dots          & $82.14$                      & $48.47$  \\
$\chi^2_{\rm det}$                                  & \dots          & $63.09$                      & $18.59$ \\
Degrees of freedom                                  & \dots          & $32$                         & $29$    \\
Reduced $\chi^2_{\rm det}$                          & \dots          & $1.97$                       & $0.64$  \\
\enddata
\tablecomments{
Parameter uncertainties correspond to the 68\% credible intervals defined by
the 16th and 84th percentiles of the marginalized posterior distributions.
The likelihood, BIC, and diagnostic $\chi^2_{\rm det}$ values are evaluated
using the maximum-likelihood solution. The diagnostic $\chi^2_{\rm det}$ is
computed using detected flux measurements only. Upper limits remain
included in the likelihood analysis but are excluded from this statistic.
For asymmetric flux uncertainties, the uncertainty corresponding to the direction
of the model deviation is adopted. The number of degrees of freedom is defined
as $\nu=N_{\rm det}-k$, where $N_{\rm det}$ is the number of detected flux
measurements and $k$ is the number of free model parameters, and the reduced
statistic is $\chi^2_{\nu,\rm det}=\chi^2_{\rm det}/\nu$. In the
lepto-hadronic model, $n_{\rm H}=20~{\rm cm^{-3}}$ is fixed. Entries marked
with $\cdots$ do not apply to the purely leptonic model. The
electron-to-proton normalization ratio $K_{\rm ep}$ is defined in
Equation~(\ref{eq:kep}).
}
\end{deluxetable*}

\subsection{Full-template Configuration and Sensitivity to the Spatial Treatment}
\label{sec:model_full}

We next repeat the modeling with the integrated full-template dataset,
which retains the emission spatially coincident with the MC-core/S0 region.
Because the GeV and TeV inputs change between the two configurations while
the radio and X-ray data are kept fixed, this comparison shows directly how
the inferred particle properties respond to the $\gamma$-ray spatial
treatment. The MC-core-excised H.E.S.S. spectrum remains an approximation
based on the rescaling described in Section~\ref{sec:configurations}, and not
an independent spatially resolved TeV measurement.

The leptonic fit to the full-template spectrum yields $\alpha_e=2.34\pm0.02$, 
$E_{\rm cut,e}=30.3^{+2.4}_{-2.1}$ TeV, $A_e=(4.99\pm0.22)\times10^{34}$ 
eV$^{-1}$, and $B=13.8\pm0.6$ $\mu$G. The corresponding total electron energy 
above 1 GeV is $W_e=(2.35^{+0.20}_{-0.19})\times10^{48}$ erg. Compared with the 
fiducial fit, the electron index
changes only slightly, the cutoff remains close to 30~TeV, and the magnetic
field decreases from about 15.1 to 13.8~$\mu$G, accompanied by an increase in
the electron normalization. These variations are comparable to the posterior
uncertainties and leave the leptonic solution qualitatively unchanged. The
magnetic field also remains below the $B\simeq27$ to $29~\mu$G of the purely
leptonic models of \citet{Guo_2018}.

\begin{figure*}[ht!]
\centering

\begin{minipage}{0.49\textwidth}
    \centering
    \includegraphics[width=\textwidth]{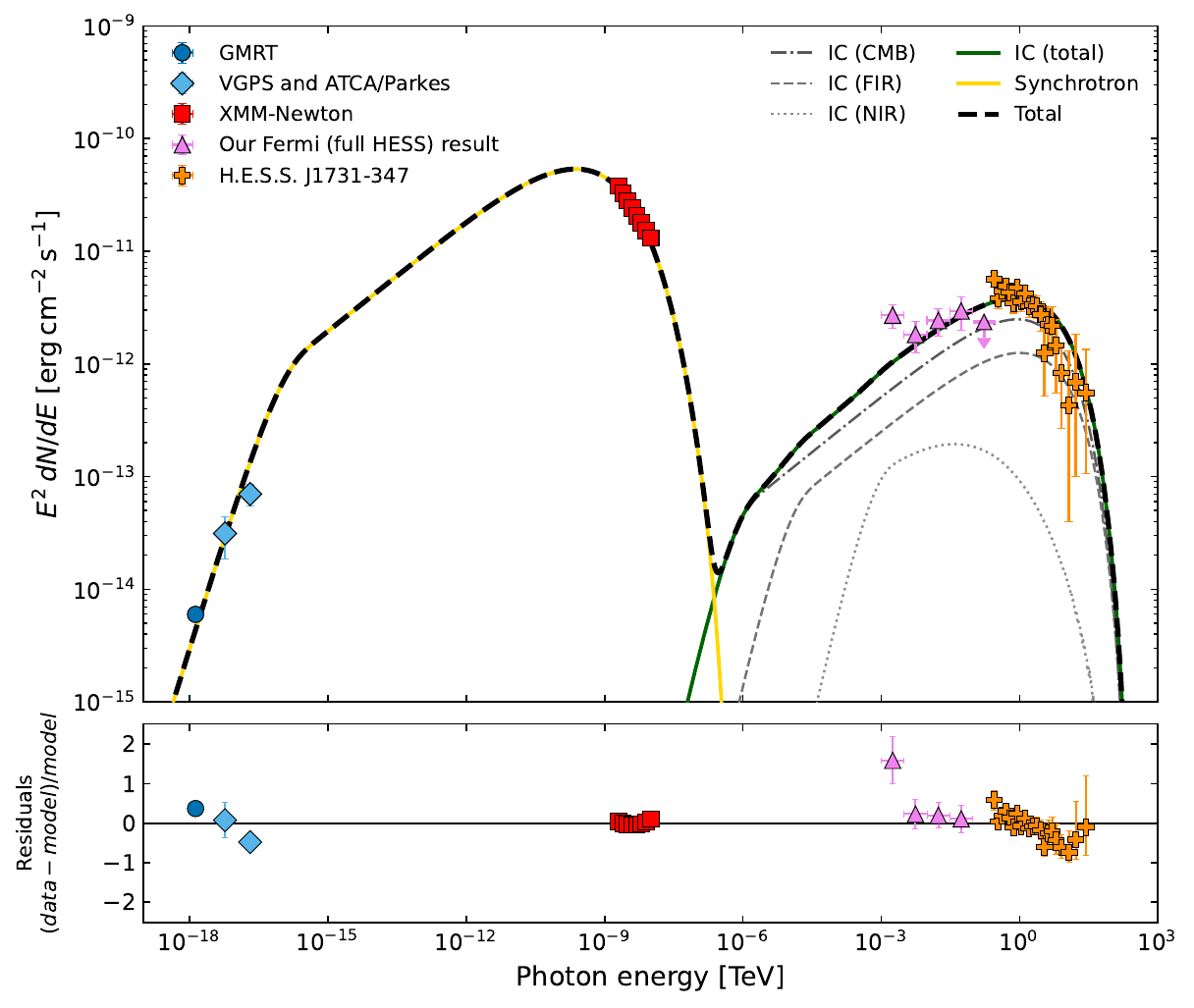}\\[0.3em]
    {\small (a) Leptonic Model}
\end{minipage}
\hfill
\begin{minipage}{0.49\textwidth}
    \centering
    \includegraphics[width=\textwidth]{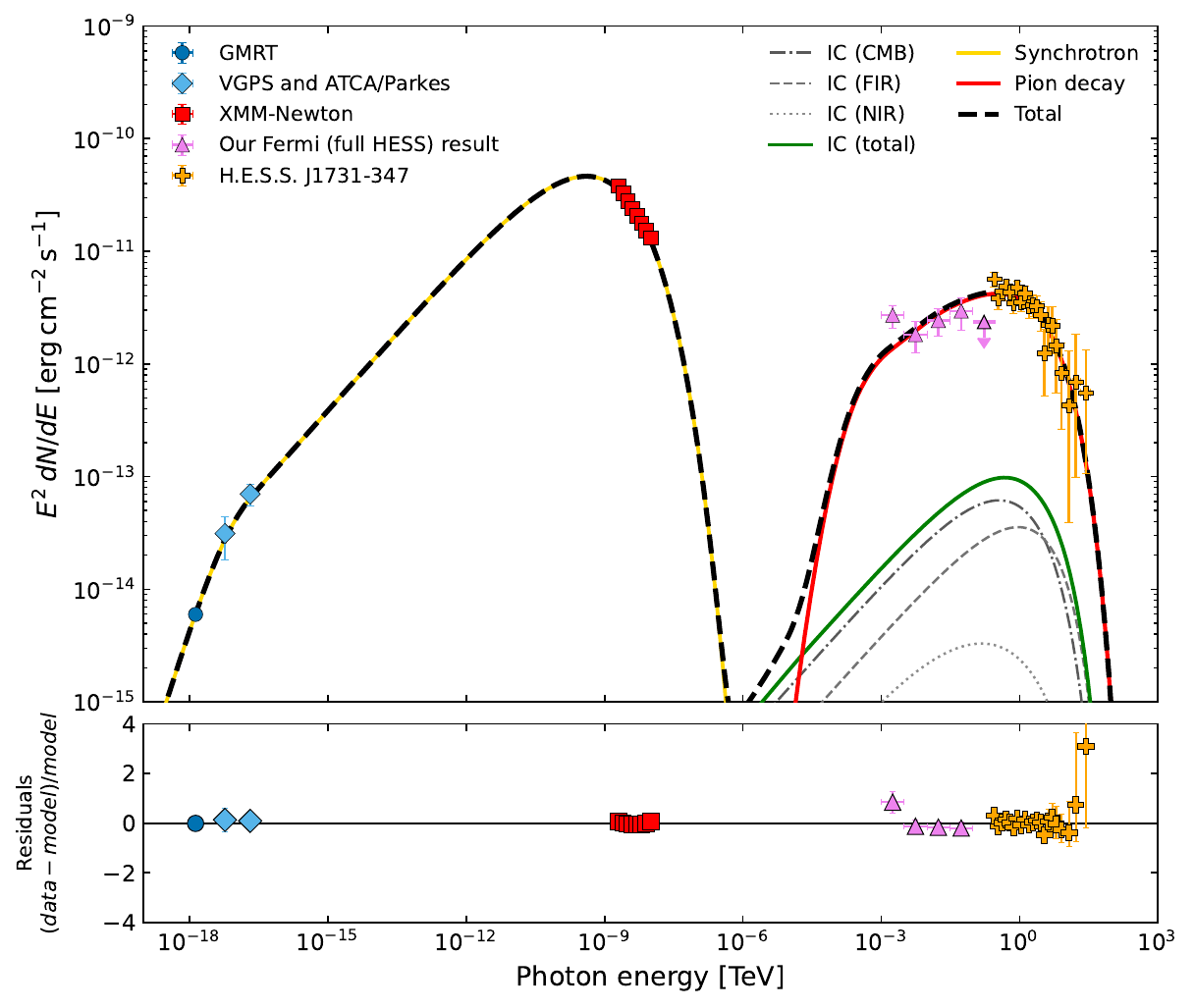}\\[0.3em]
    {\small (b) Lepto-hadronic Model}
\end{minipage}

\caption{Broadband spectral energy distributions of HESS~J1731$-$347 for the
integrated full-template configuration. Panel (a) shows the purely
leptonic model, with synchrotron and IC emission, while
panel (b) shows the lepto-hadronic model including an additional
$\pi^0$-decay component from proton-proton interactions. The
individual IC contributions from the CMB, FIR, and NIR photon fields
are shown together with their sum. The lower panels show the fractional
residuals, $(F_{\rm data}-F_{\rm model})/F_{\rm model}$.}
\label{fig:sed_full}
\end{figure*}

The lepto-hadronic fit yields $\alpha_e=2.08\pm0.03$, 
$E_{\rm cut,e}=11.4^{+1.8}_{-1.5}$ TeV, $A_e=(1.3^{+0.7}_{-0.4})\times10^{33}$ eV$^{-1}$, 
and $B=83.4^{+25.5}_{-21.7}$ $\mu$G, with $K_{\rm ep}=(2.6^{+1.5}_{-0.9})\times10^{-3}$ 
and a proton normalization of $A_p=(4.92\pm0.48)\times10^{35}$ eV$^{-1}$. The 
proton population has $\alpha_p=1.81\pm0.08$ and 
$E_{\rm cut,p}=32.2^{+13.6}_{-9.6}$ TeV. The corresponding total energies 
above 1 GeV are $W_e=(2.28^{+1.00}_{-0.61})\times10^{46}$ erg and 
$W_p=(6.28^{+0.79}_{-0.73})\times10^{48}$ erg. The electron component and the 
magnetic field are therefore stable, with $\alpha_e\simeq2.08$,
$E_{\rm cut,e}\simeq11$~TeV, and $B\simeq83$ to $85~\mu$G in both
configurations, and the field remains similar to the hybrid solutions of
\citet{Guo_2018}. The proton parameters respond more noticeably. Relative to
the fiducial fit, the full-template solution has a softer proton index
($1.81$ versus $1.69$), a higher cutoff energy ($32.2$ versus $23.5$~TeV),
and a larger proton normalization. These changes accompany the larger GeV
flux of the full-template configuration toward the lower-energy end of the
LAT spectrum, where the contribution associated with S0 is greatest
(Section~\ref{sec:configurations}). The credible intervals of the proton
parameters nevertheless remain broadly compatible, so we do not interpret
these shifts as evidence for distinct proton populations, nor as a
measurement of the particle population within the MC-core itself. They show
instead that the inferred proton component is more sensitive than the
electron component to the adopted treatment of the $\gamma$-ray emission.
The broadband decompositions are shown in Figure~\ref{fig:sed_full}.

For the full-template dataset, the leptonic model yields
$\ln\hat{\mathcal{L}}=-38.05$ and ${\rm BIC}=90.54$, whereas the
lepto-hadronic model yields
$\ln\hat{\mathcal{L}}=-13.66$ and ${\rm BIC}=52.59$, giving
\begin{equation}
\Delta{\rm BIC}_{\rm full}
=
{\rm BIC}_{\rm lep}
-
{\rm BIC}_{\rm LH}
=
37.95.
\end{equation}
This value again corresponds to very strong evidence for the
lepto-hadronic model on the scale of \citet{Kass1995}. The complementary
diagnostic gives the same ordering, with $\chi^2_{\rm det}=71.49$ for 32
degrees of freedom ($\chi^2_{\nu,{\rm det}}=2.23$) for the leptonic solution
and $\chi^2_{\rm det}=22.71$ for 29 degrees of freedom
($\chi^2_{\nu,{\rm det}}=0.78$) for the lepto-hadronic one. The parameters
and statistical indicators are summarized in Table~\ref{tab:params_full}.

The relative ordering of the two emission scenarios is therefore
preserved under both spatial treatments. In each configuration the
lepto-hadronic model yields the lower BIC and the better complementary
goodness-of-fit diagnostic. Within the adopted one-zone framework and
environmental assumptions, the preference for the lepto-hadronic solution is
not driven solely by the treatment of the S0/MC-core region.

\begin{deluxetable*}{llcc}[ht!]
\tablecaption{Posterior parameter estimates and statistical indicators
for the full-template configuration of HESS~J1731$-$347,
considering the purely leptonic and lepto-hadronic emission models.
\label{tab:params_full}}
\tablecolumns{4}
\tablehead{
\colhead{Parameter} & \colhead{Unit} &
\colhead{Leptonic} & \colhead{Lepto-hadronic}
}
\startdata
\noalign{\smallskip}
\multicolumn{4}{c}{Electron Distribution} \\
\noalign{\smallskip} \tableline \noalign{\smallskip}
Spectral index $\alpha_e$                           & \dots          & $2.34 \pm 0.02$               & $2.08 \pm 0.03$ \\
Cutoff energy $E_{\rm cut,e}$                       & TeV            & $30.3\,^{+2.4}_{-2.1}$        & $11.4^{+1.8}_{-1.5}$ \\
Normalization $A_e$                                 & eV$^{-1}$      & $(4.99\pm 0.22) \times 10^{34}$  & $(1.3^{+0.7}_{-0.4})\times10^{33}$ \\
Total electron energy $W_e$ ($E > 1$~GeV)           & erg            & $(2.35^{+0.20}_{-0.19})\times10^{48}$          & $(2.28^{+1.00}_{-0.61})\times10^{46}$ \\
\noalign{\smallskip} \tableline \noalign{\smallskip}
\multicolumn{4}{c}{Proton Distribution} \\
\noalign{\smallskip} \tableline \noalign{\smallskip}
Spectral index $\alpha_p$                           & \dots          & \nodata                       & $1.81\pm 0.08$ \\
Cutoff energy $E_{\rm cut,p}$                       & TeV            & \nodata                       & $32.2\,^{+13.6}_{-9.6}$ \\
Normalization $A_p$                                 & eV$^{-1}$      & \nodata                       & $(4.92 \pm 0.48) \times 10^{35}$\\
Total proton energy $W_p$ ($E > 1$~GeV)             & erg            & \nodata                       & $(6.28^{+0.79}_{-0.73})\times10^{48}$ \\
Electron-to-proton ratio $K_{\rm ep}$               & \dots          & \nodata                       & $(2.6^{+1.5}_{-0.9})\times10^{-3}$ \\
\noalign{\smallskip} \tableline \noalign{\smallskip}
\multicolumn{4}{c}{Environmental Conditions} \\
\noalign{\smallskip} \tableline \noalign{\smallskip}
Magnetic field $B$                                  & $\mu$G         & $13.8\pm 0.6$                & $83.4 \, ^{+25.5}_{-21.7}$ \\
Distance $d$                                        & kpc            & $3.2$ (fixed)                & $3.2$ (fixed) \\
\noalign{\smallskip} \tableline \noalign{\smallskip}
\multicolumn{4}{c}{Statistical Indicators} \\
\noalign{\smallskip} \tableline \noalign{\smallskip}
Log-likelihood $\ln\mathcal{L}$                     & \dots          & $-38.05$                     & $-13.66$ \\
Bayesian Information Criterion, BIC                 & \dots          & $90.54$                      & $52.59$  \\
$\chi^2_{\rm det}$                                  & \dots          & $71.49$                      & $22.71$ \\
Degrees of freedom                                  & \dots          & $32$                         & $29$    \\
Reduced $\chi^2_{\rm det}$                          & \dots          & $2.23$                       & $0.78$  \\
\enddata
\tablecomments{
Same conventions as in Table~\ref{tab:params_no_mc}.
}
\end{deluxetable*}

% ============================================================
\section{Prospects for CTAO Observations}
\label{sec:ctao}
% ============================================================

\subsection{Simulation Setup}
\label{sec:ctao_setup}

To evaluate how well future observations can test the emission scenarios
examined in this work, we simulated the response of CTAO
\citep{2019scta.book.....C}. CTAO is the next-generation VHE $\gamma$-ray
observatory and will improve
substantially on current instruments in sensitivity, energy coverage, and
angular resolution. It is designed to probe non-thermal processes across a
wide range of astrophysical sources, including sites of efficient particle
acceleration. The observatory comprises two sites, one in the Northern
Hemisphere at La Palma and one in the Southern Hemisphere near Paranal,
providing nearly full-sky coverage.
The instrument layout includes three telescope classes (large, medium, and
small), each optimized for a specific energy regime. This configuration
yields continuous energy coverage from a few tens of GeV to beyond
$100$~TeV. Compared to current facilities, CTAO will offer better
sensitivity, finer angular resolution, better energy reconstruction, and a
larger field of view. These improvements are essential for resolving complex
source morphologies, detecting spectral cutoffs or breaks, and identifying
the physical mechanisms behind CR acceleration
\citep{2019scta.book.....C}.

Before performing detailed simulations, we examined the annual visibility of
HESS~J1731$-$347 (RA~$=263.01^\circ$, Dec~$=-34.75^\circ$) for both CTAO
sites as a function of zenith angle ($z$). Table~\ref{tab:visibility} lists the annual visibility durations for the southern
and northern CTAO sites in three zenith-angle intervals:
$[10^\circ, 30^\circ)$, $[30^\circ, 50^\circ)$, and
$[50^\circ, 70^\circ)$, represented by their central values
of $20^\circ$, $40^\circ$, and $60^\circ$, respectively.
At CTAO South, HESS~J1731$-$347 reaches a minimum zenith
angle of approximately $10^\circ$ and therefore does not
pass directly through the zenith. The source is only accessible from the 
southern site at low zenith angles ($20^\circ$ and $40^\circ$), where the 
atmospheric transmission and instrument response are optimal. At 
$z=60^\circ$, the northern array becomes partially visible (757.5~h), but 
observations at such
large zenith angles suffer from increased energy thresholds and degraded
angular resolution. The southern array at $z=20^\circ$ offers the cleanest
observing window, with 797.5~h per year. All subsequent simulations are
therefore based on the full southern array at $z=20^\circ$, with an
exposure time of $t_{\mathrm{obs}} = 50$~h, which is only a fraction of the
total available time and can be scheduled flexibly.

\setlength{\tabcolsep}{15pt}
\begin{table}
\centering
\renewcommand{\arraystretch}{1.3}
\caption{Annual visibility durations (in hours) for
HESS~J1731$-$347 at the CTAO South and North sites.
Values are based on the source coordinates
(RA = $263.01^\circ$, Dec = $-34.75^\circ$).
The zenith angles of $20^\circ$, $40^\circ$,
and $60^\circ$ correspond to the intervals
$[10^\circ,30^\circ)$, $[30^\circ,50^\circ)$,
and $[50^\circ,70^\circ)$, respectively.
At CTAO South, the source reaches a minimum
zenith angle of approximately $10^\circ$
and therefore does not transit through the zenith.}
\label{tab:visibility}
\begin{tabular}{lccc}
\hline
\hline
\textbf{Array} & \multicolumn{3}{c}{\textbf{Zenith Angle}} \\
\cline{2-4}
& $20^\circ$ & $40^\circ$ & $60^\circ$ \\
\hline
CTAO South & 797.5 & 594.0 & 604.5 \\
CTAO North & 0.0 & 0.0 & 757.5 \\
\hline
\hline
\end{tabular}
\vspace{0.2cm}
\end{table}

We performed dedicated three-dimensional simulations of the expected
$\gamma$-ray emission and derived flux points using the \texttt{Gammapy}
package \citep{2023A&A...678A.157D, gammapy}. The simulations adopt the
instrument response functions (IRFs) of the Beta configuration, which
represents the current deployment stage of the observatory \citep{cta_observatory_2026_22871179}.
We used the IRFs of the southern array evaluated at a zenith angle of
$z = 20^\circ$ under dark-sky conditions, covering the energy range from 50~GeV to 100~TeV. This band covers the nominal performance range of the southern site, which includes all three telescope classes, and the chosen zenith angle is well matched to the declination of the source, which gives the best sensitivity and instrumental performance.

The $\gamma$-ray emission was simulated for the purely leptonic and
lepto-hadronic best-fit models of Section~\ref{sec:scenarios}, each applied
to both the full-template and the MC-core-excised configurations. In all
cases, the observed H.E.S.S. flux map was adopted as the spatial template
\citep{HESS2011,2018A&A...612A...1H}. For the MC-core-excised configuration,
a discrete spatial mask was applied to the map to remove the flux originating
from that region. Within \texttt{Gammapy}, this two-dimensional array is
implemented as a normalized spatial model. The map is integrated and each
pixel is divided by the total integral, which converts it into a spatial
probability density function (PDF) with units of $\mathrm{sr}^{-1}$. This
procedure decouples the spatial morphology of the source from its spectral
energy distribution. The spectral models therefore set the total integrated
flux, while the normalized template acts as a two-dimensional weighting
function that distributes the simulated photons across the field of view
according to the observed morphology.

For each emission scenario, we generated $3000$ independent simulated
realizations by introducing Poisson fluctuations into both the expected
source signal and the instrumental background, as defined by the adopted
IRFs. The significance maps and reconstructed spectral flux points were then
derived through a standard binned Poisson likelihood maximization.

\subsection{Simulated Detection, Morphology, and Spectra}
\label{sec:ctao_results}

The simulated datasets show that CTAO will detect the $\gamma$-ray emission
with high statistical significance. For a total exposure of
$t_{\mathrm{obs}} = 50$~h, the leptonic model yields detection significances
of approximately $13\sigma$ and $16\sigma$ for the MC-core-excised and full spatial templates, respectively. Under the lepto-hadronic model, these
values reach approximately $15\sigma$ and $18\sigma$. The source would
therefore be robustly detected in either scenario, with the photon statistics
needed for detailed spectral and morphological characterization.

\begin{figure*}
    \centering
    \gridline{
        \fig{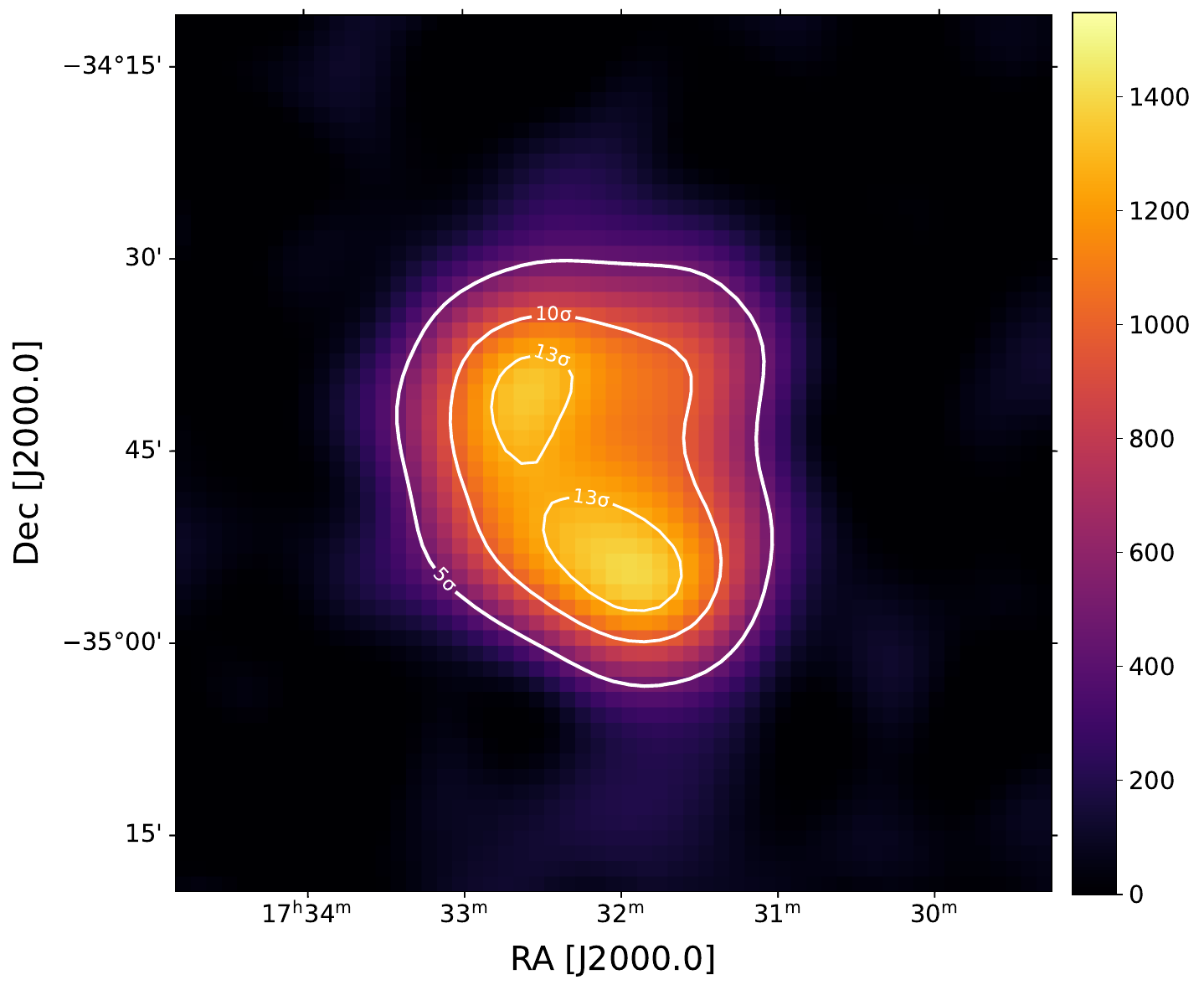}{0.49\textwidth}{}
        \fig{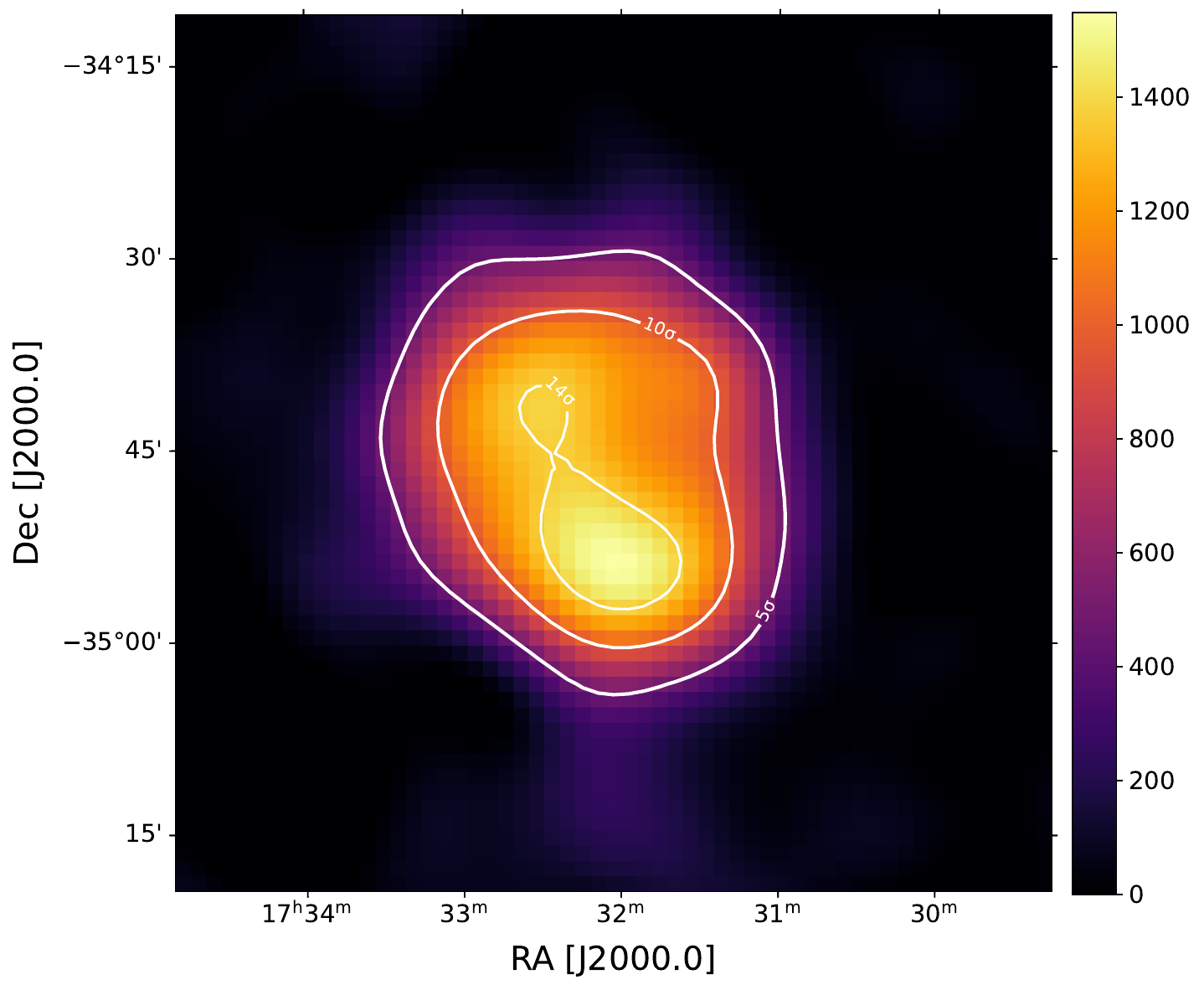}{0.49\textwidth}{}
    }
    \gridline{
        \fig{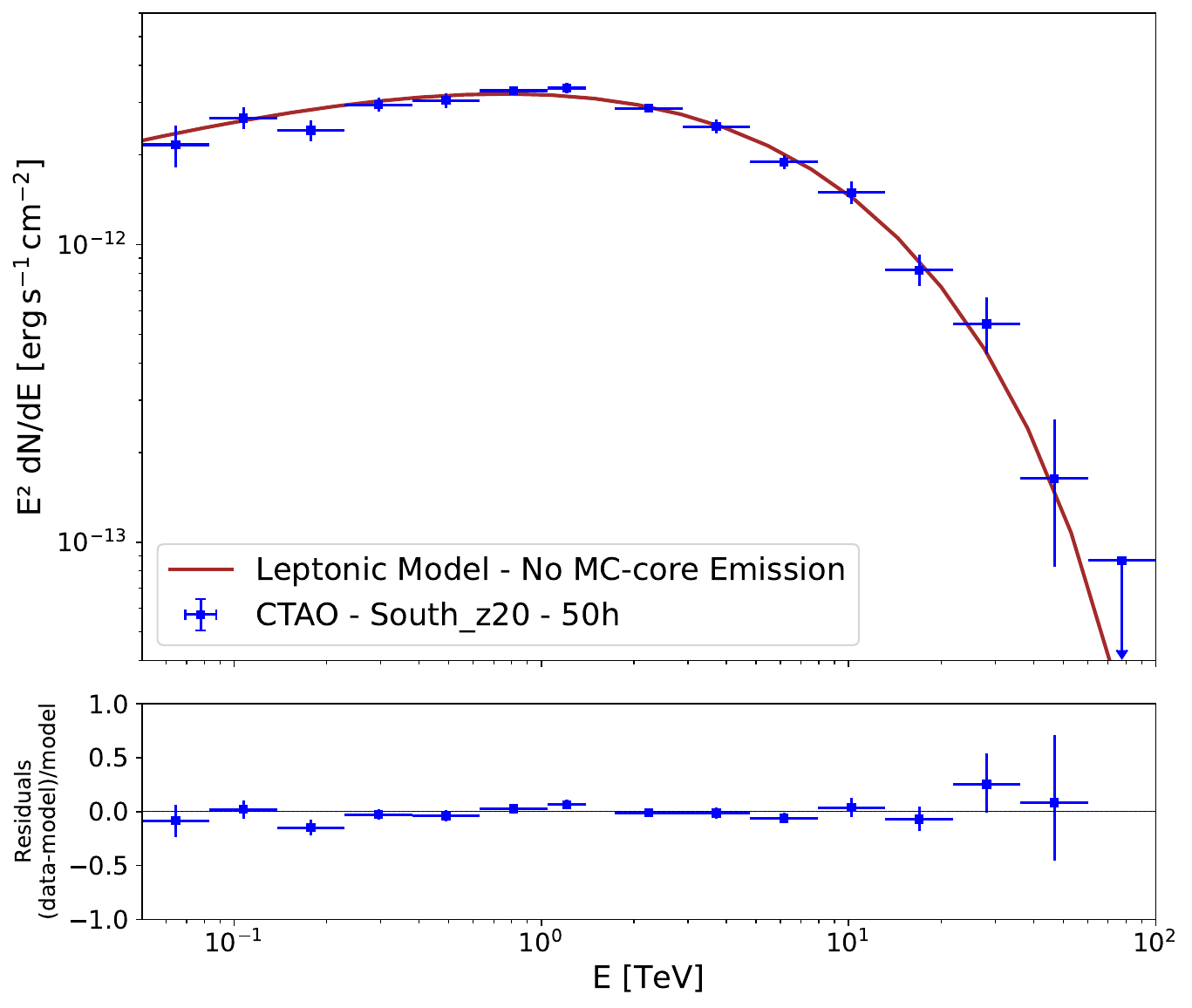}{0.49\textwidth}{(a) MC-core-excised emission, leptonic model}
        \fig{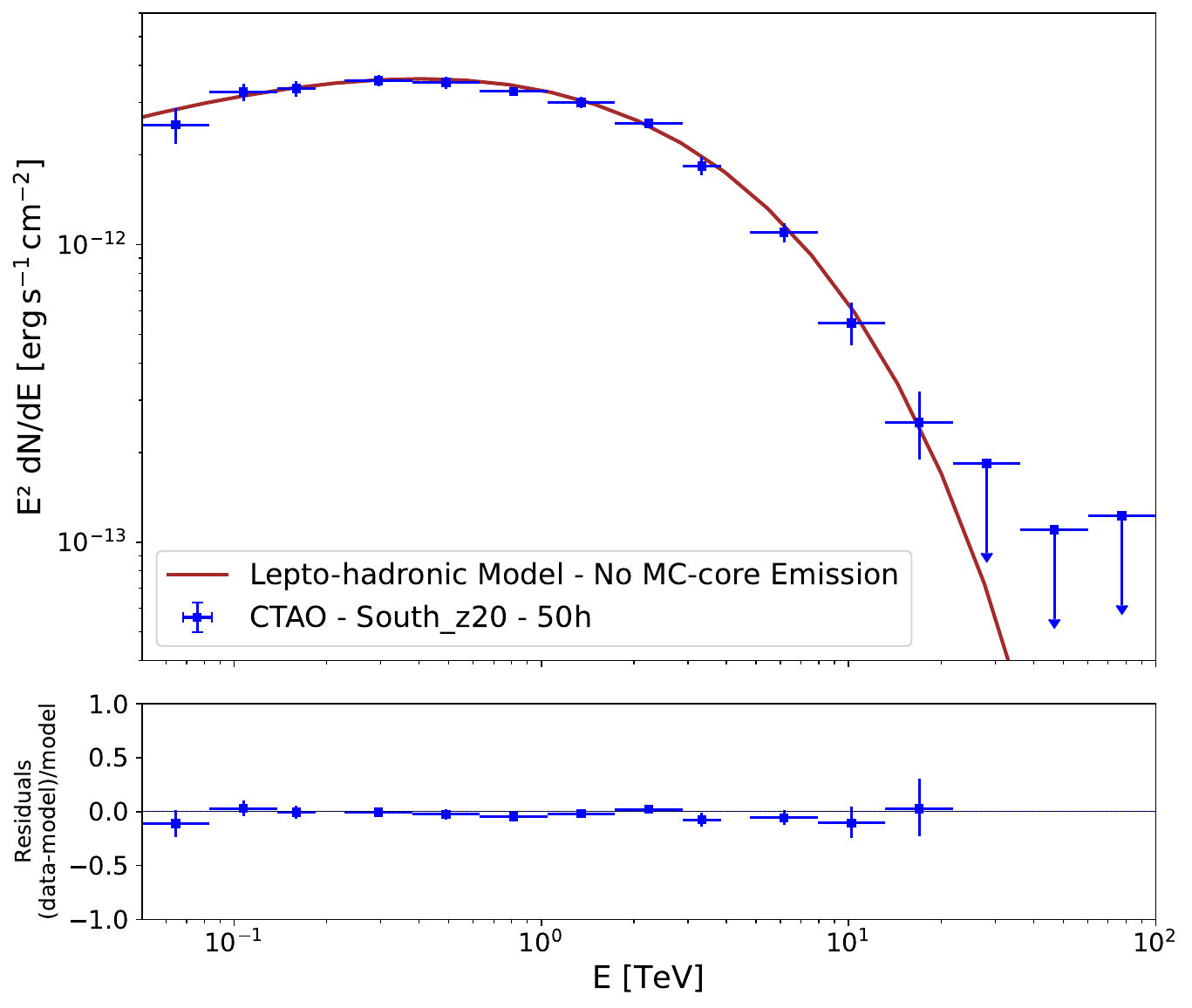}{0.49\textwidth}{(b) MC-core-excised emission, lepto-hadronic model}
    }
\caption{Simulated CTAO observations of HESS~J1731$-$347 for the
MC-core-excised configuration, for an exposure time of
$t_{\mathrm{obs}} = 50$~h with the full southern array IRFs at a zenith angle
of $z = 20^{\circ}$ under dark-sky conditions. The upper panels show the simulated VHE $\gamma$-ray excess maps for the leptonic (left) and lepto-hadronic (right) scenarios. The maps use a spatial binning of $0.02^{\circ}$ and are smoothed with a Gaussian kernel of $0.04^{\circ}$. Solid white contours indicate significance levels of $5\sigma$, $10\sigma$, $13\sigma$, and $14\sigma$. The color scale gives the simulated excess counts per pixel.
The lower panels show the reconstructed flux points (markers) together with
the input SED model of each scenario (solid lines). The simulated points
recover the input models. The two best-fit spectra differ in curvature below
a few TeV, where the flux points are most precise, and in their decline at
the highest energies.}
    \label{fig:ctao_simul_mc-core}
\end{figure*}

\begin{figure*}
    \centering
    \gridline{
        \fig{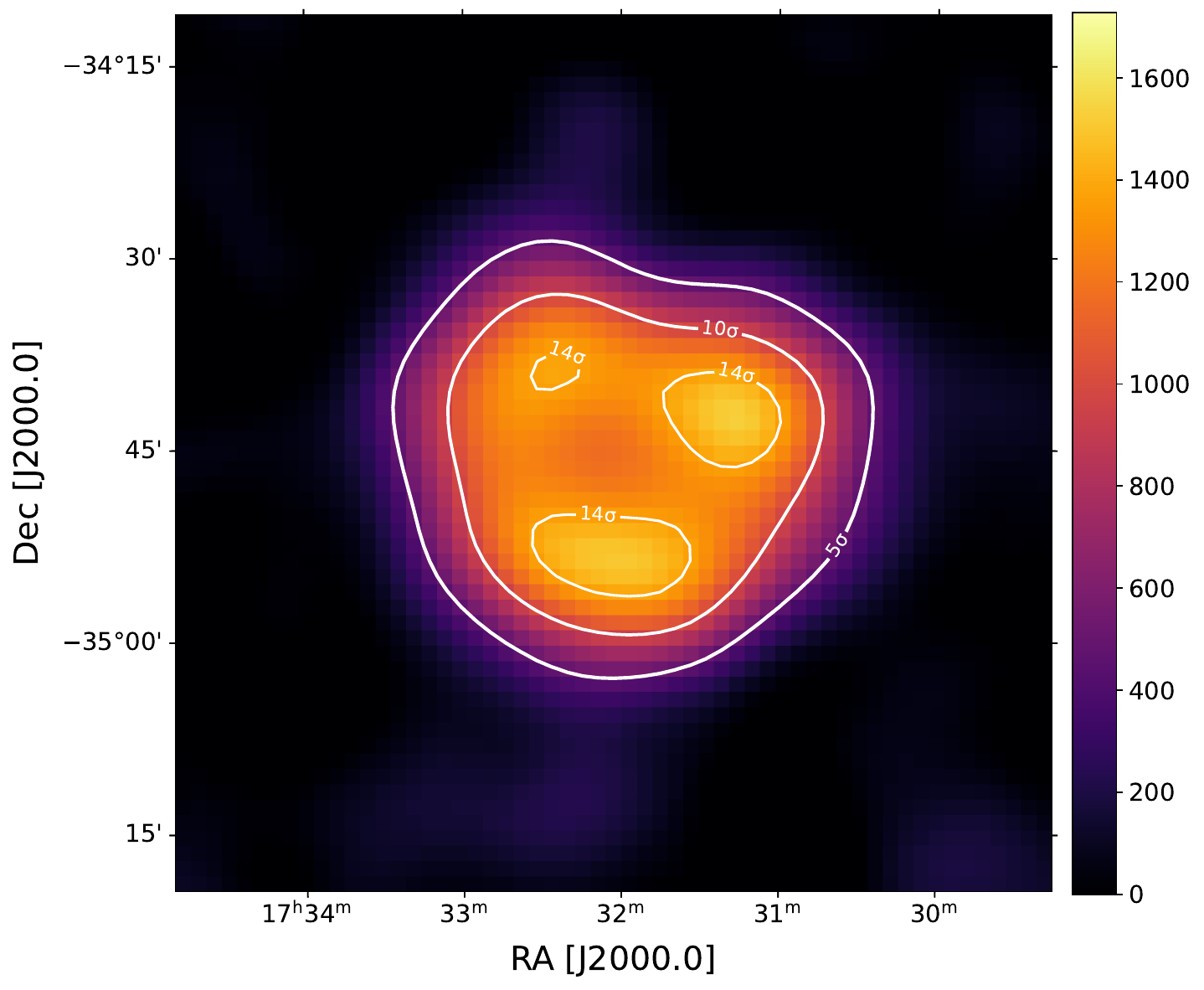}{0.49\textwidth}{}
        \fig{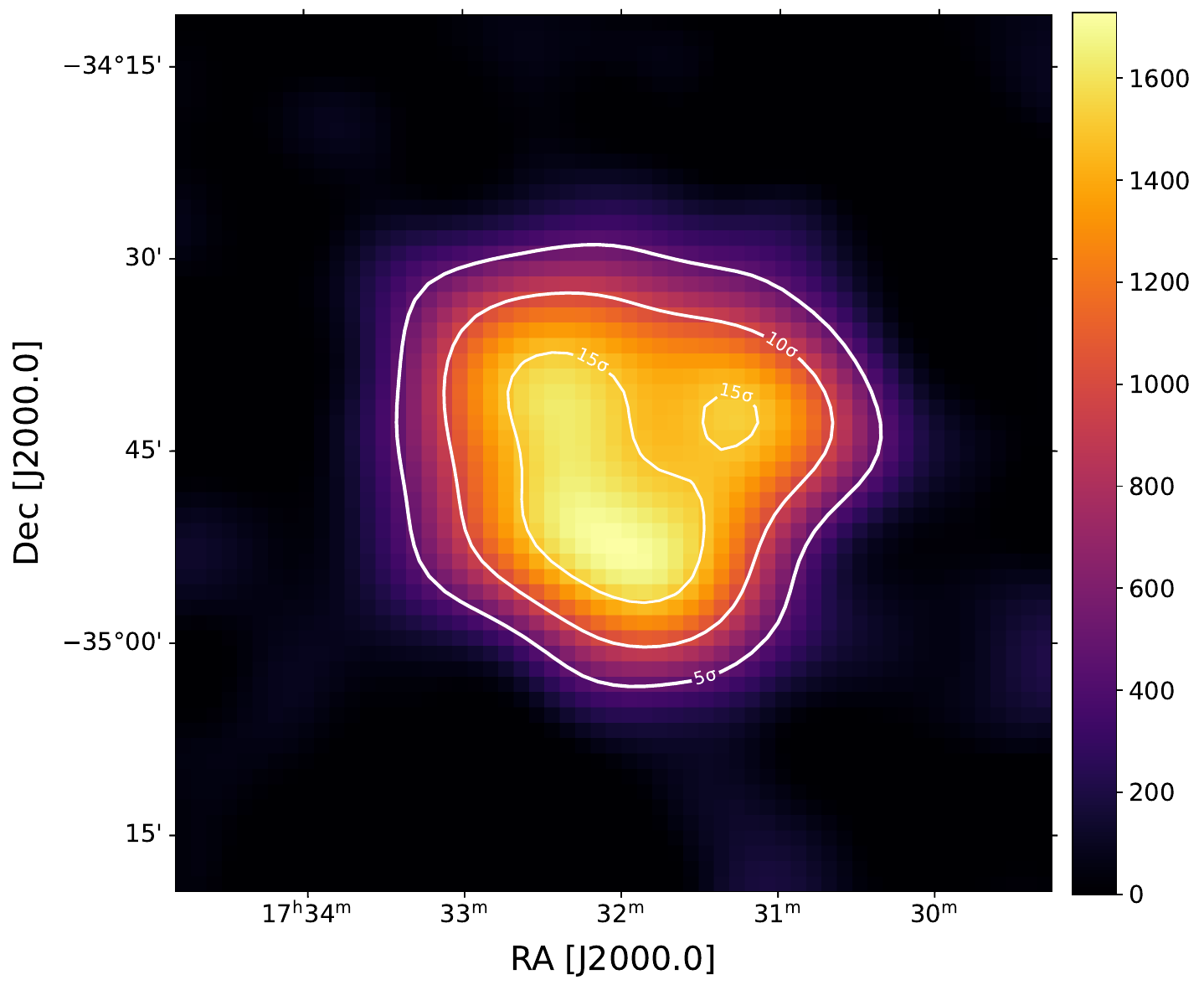}{0.49\textwidth}{}
    }
    \gridline{
        \fig{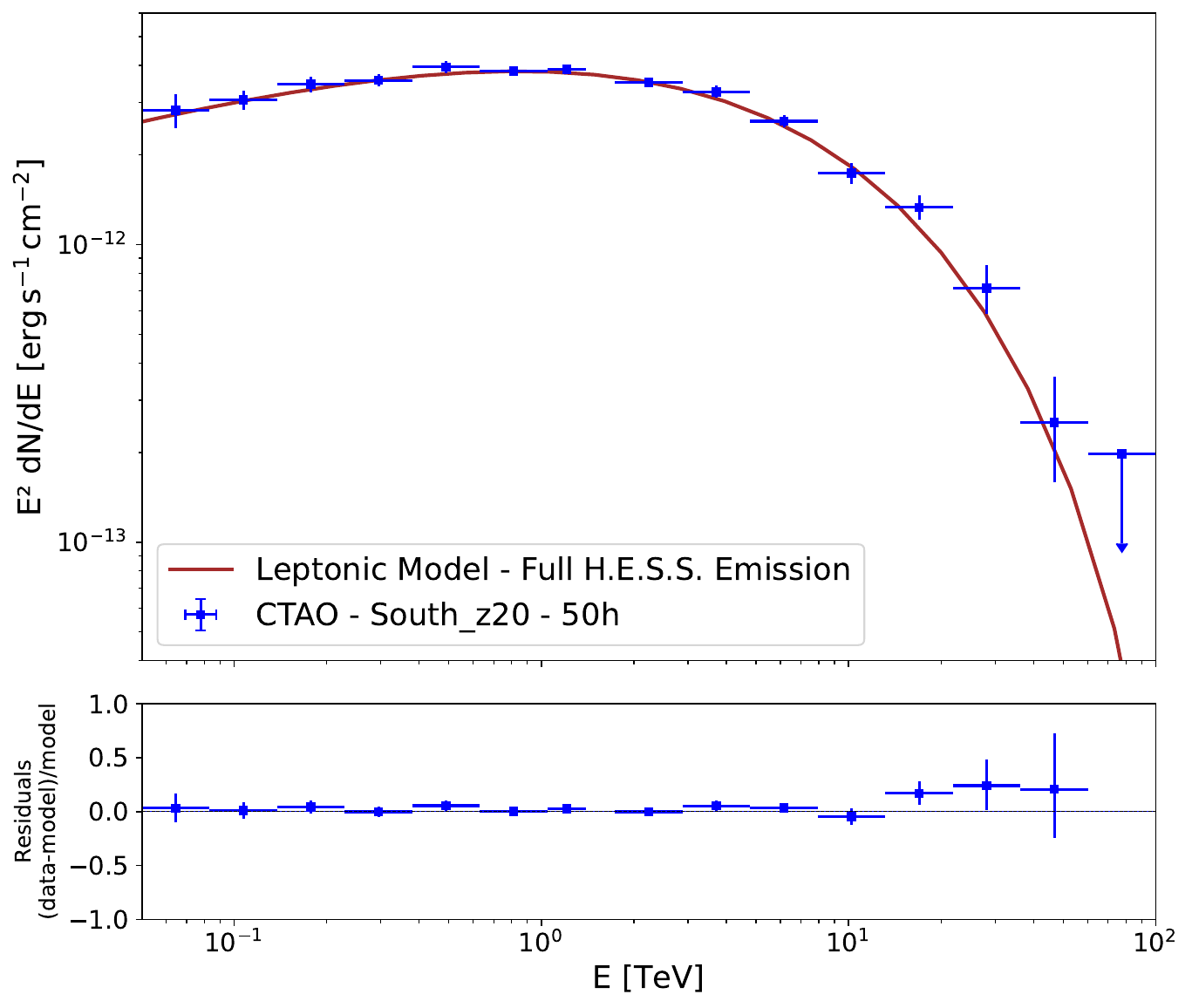}{0.49\textwidth}{(a) Full H.E.S.S. emission, leptonic model}
        \fig{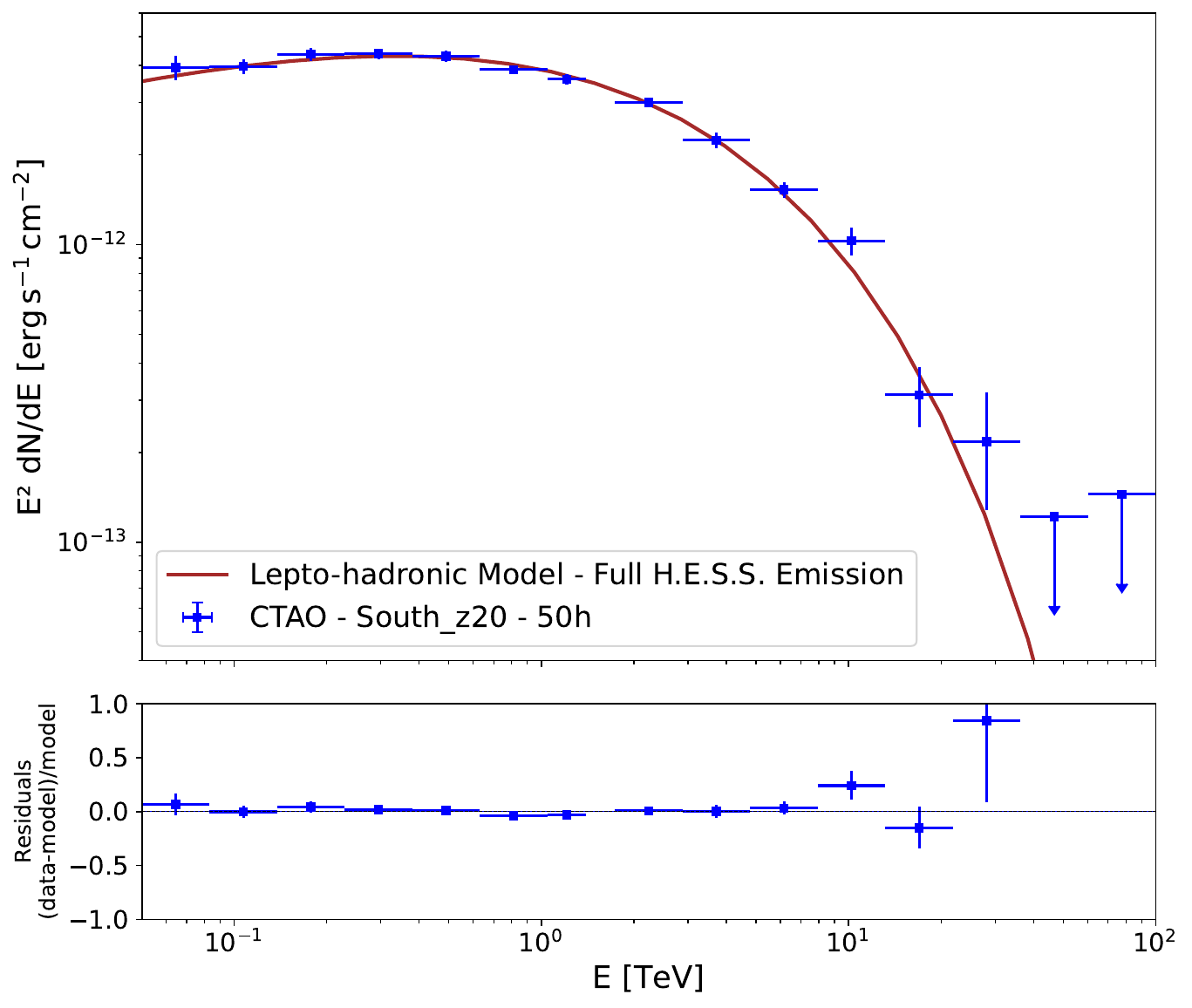}{0.49\textwidth}{(b) Full H.E.S.S. emission, lepto-hadronic model}
    }
\caption{Same as Figure~\ref{fig:ctao_simul_mc-core}, but for the
integrated full-template configuration.}
    \label{fig:ctao_simul_full}
\end{figure*}

Figure~\ref{fig:ctao_simul_mc-core} presents the simulated excess maps
and the reconstructed SEDs for the MC-core-excised configuration. The
significance contours in the upper panels show that CTAO will resolve the
shell morphology of HESS~J1731$-$347, owing to its angular resolution of
$\sim0.05^\circ$ at TeV energies, which is substantially better than that of
current-generation instruments. In the leptonic case, the emission appears
more diffuse, with lower excess peaks that are separated and located toward
the periphery of the remnant. The lepto-hadronic case shows a more
concentrated and connected structure with higher significance levels.
Because both scenarios use the same normalized spatial template, these
differences do not represent an intrinsic difference in source morphology.
They follow from the different spectral shapes of the two models. As the lower
panels show, the lepto-hadronic model predicts a higher flux below
$\sim1$~TeV, where most of the simulated photons are collected, which raises
the excess and the significance across the whole template.

The reconstructed flux points in the lower panels show that CTAO can place meaningful constraints on the shape of the two best-fit spectra, although a definitive discrimination between the scenarios is not possible from these simulations alone. The most precise part of the simulated spectrum lies between a few hundred GeV and a few TeV, where the flux points have the smallest uncertainties. Furthermore, the inclusion of two LSTs at the Southern site under the Beta configuration not only extends the lower energy limit down to $\sim50$~GeV, but also significantly reduces the uncertainties in this critical sub-TeV to TeV regime. In this energy band, the leptonic spectrum is nearly flat in $E^2\,dN/dE$, whereas the lepto-hadronic spectrum has a higher flux at the lowest energies and is already curving downward. This difference in
curvature over the best-measured part of the CTAO band is likely more
constraining than the behavior at tens of TeV, where the flux points have
larger uncertainties and some bins provide only upper limits.

At higher energies, the lepto-hadronic spectrum steepens markedly above
$\sim5$~TeV. Because $\pi^0$-decay photons carry on average about one tenth of
the energy of the parent proton, the proton cutoff at
$E_{\rm cut,p}\simeq23.5$~TeV produces a $\gamma$-ray rollover at a few TeV
and not at the proton cutoff energy itself. The purely leptonic model retains
a harder tail, associated with the electron cutoff at
$E_{\rm cut,e}\simeq29$~TeV. For these specific best-fit models, the
simulated flux points also separate the two predictions in the 10 to 100~TeV
range.

The full-template configuration (Figure~\ref{fig:ctao_simul_full})
shows similar characteristics. The morphological differences between the two
spatial configurations arise from the normalization of the spatial model.
The additional flux of the full-template scenario is not confined to the
region where the MC-core was previously excluded, but is redistributed over
the entire area covered by the H.E.S.S. template. Even with this
redistribution, the lepto-hadronic model again shows a higher significance
peak and a more connected structure, while the leptonic emission appears in
slightly more separated regions.

For both spatial configurations, the reconstructed flux points follow the
input spectral models, with fractional residuals below $0.3$ over most of the energy range. At energies above $\sim20$~TeV, the residuals approach or exceed $0.5$, as expected from the low photon statistics imposed by the exponential cutoffs in the primary particle distributions. CTAO can therefore recover the intrinsic spectral properties of the source for a given input model. A more precise TeV spectrum would narrow the range of particle distributions compatible with the data, but TeV spectral information alone is not expected to remove the degeneracy between leptonic and hadronic scenarios. Even for RX~J1713.7$-$3946, whose TeV flux is about an order of magnitude higher than that of HESS~J1731$-$347, the radiating particle population cannot be identified from the TeV spectrum alone \citep{HESS2018RXJ1713}. Different
combinations of parameters within each scenario can produce similar TeV
spectral shapes, and the uncertainties of the fitted parameters, which are
not propagated in the present simulations, could make the predictions of the
two models overlap.

\begin{figure*}[ht!]
\centering
\begin{minipage}{0.49\textwidth}
    \centering
    \includegraphics[width=\textwidth]{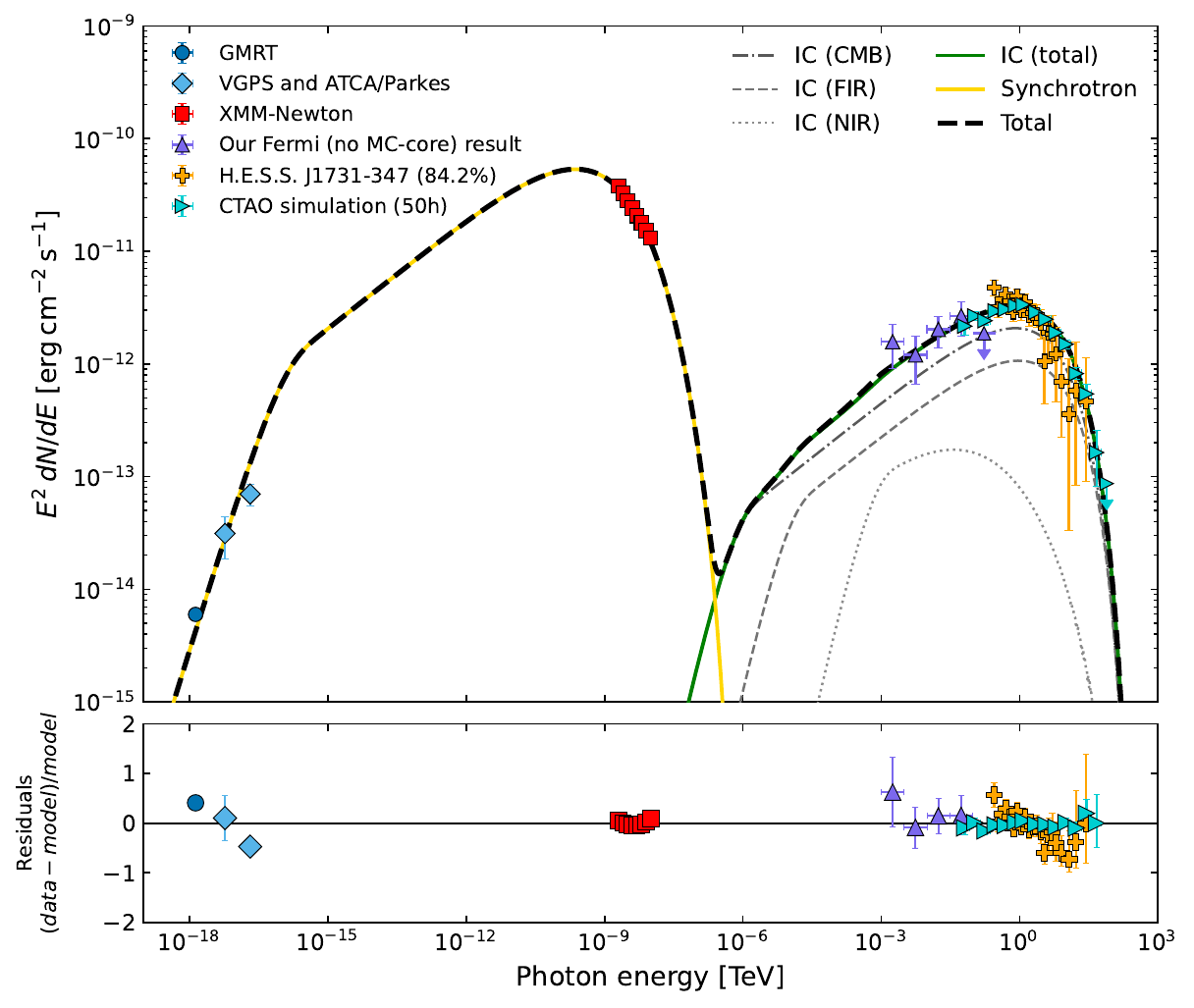}\\[0.3em]
    {\small (a) Leptonic scenario (MC-core excised)}
\end{minipage}
\hfill
\begin{minipage}{0.49\textwidth}
    \centering
    \includegraphics[width=\textwidth]{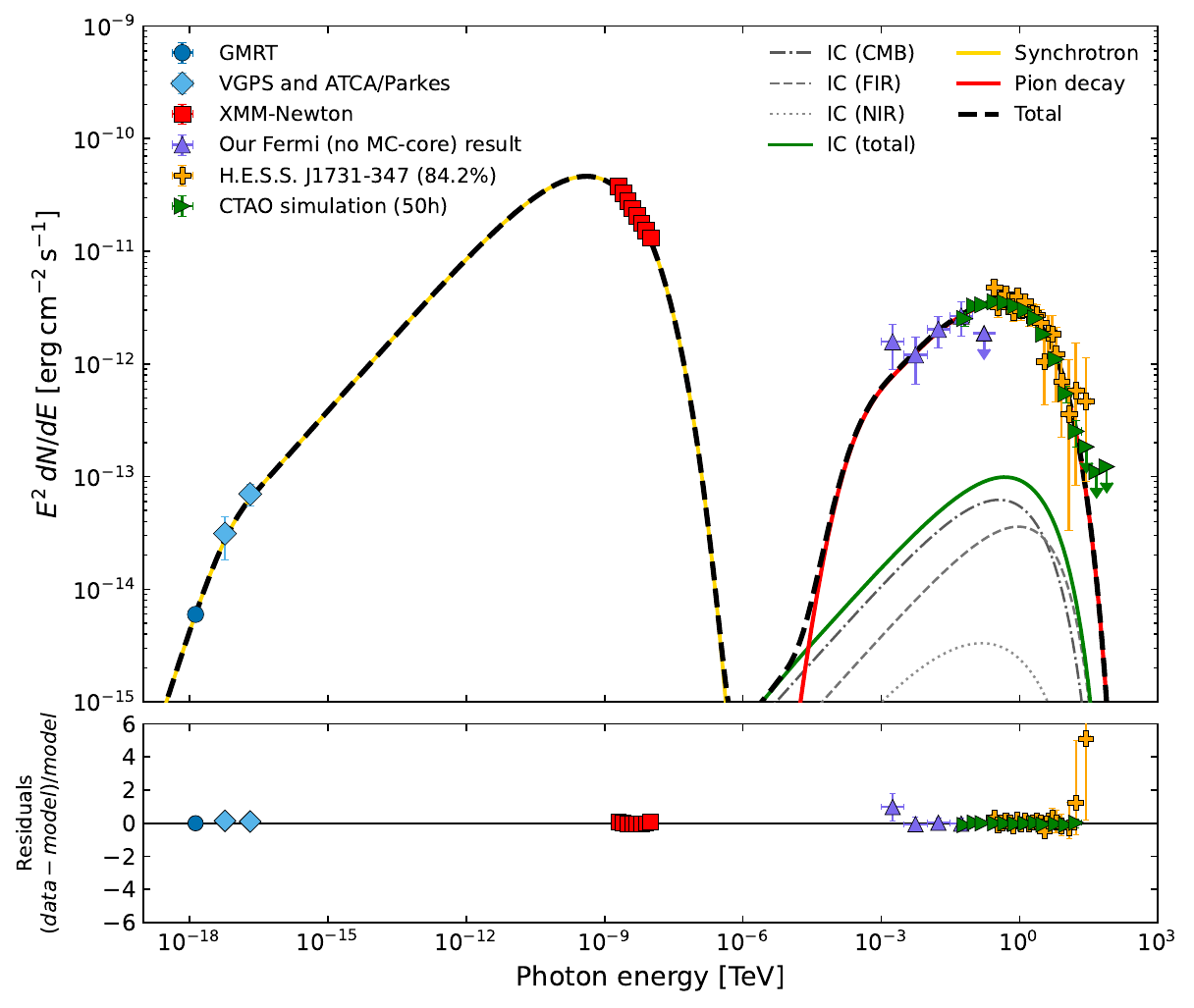}\\[0.3em]
    {\small (b) Lepto-hadronic scenario (MC-core excised)}
\end{minipage}
\begin{minipage}{0.49\textwidth}
    \centering
    \includegraphics[width=\textwidth]{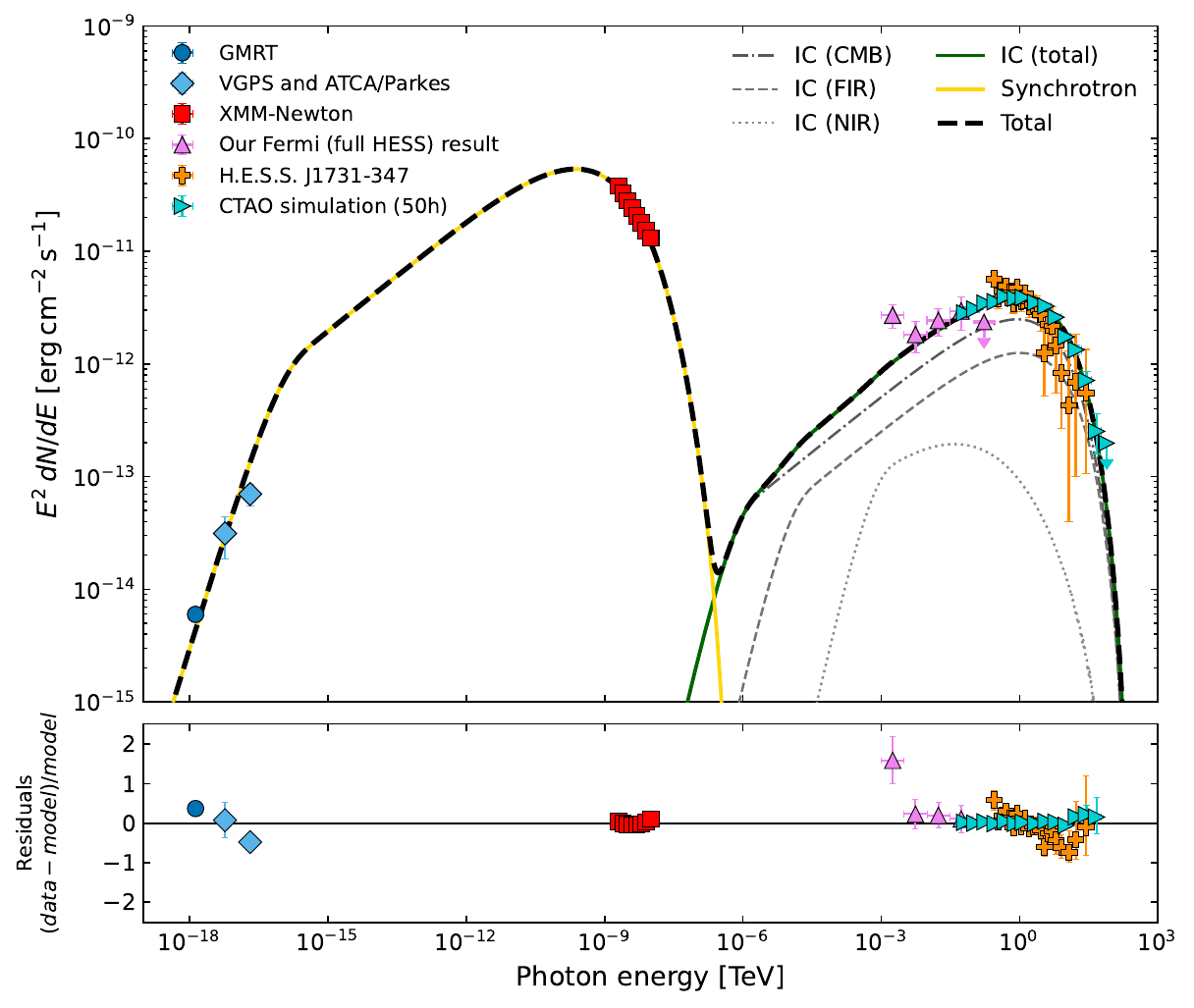}\\[0.3em]
    {\small (c) Leptonic scenario (full H.E.S.S. template)}
\end{minipage}
\hfill
\begin{minipage}{0.49\textwidth}
    \centering
    \includegraphics[width=\textwidth]{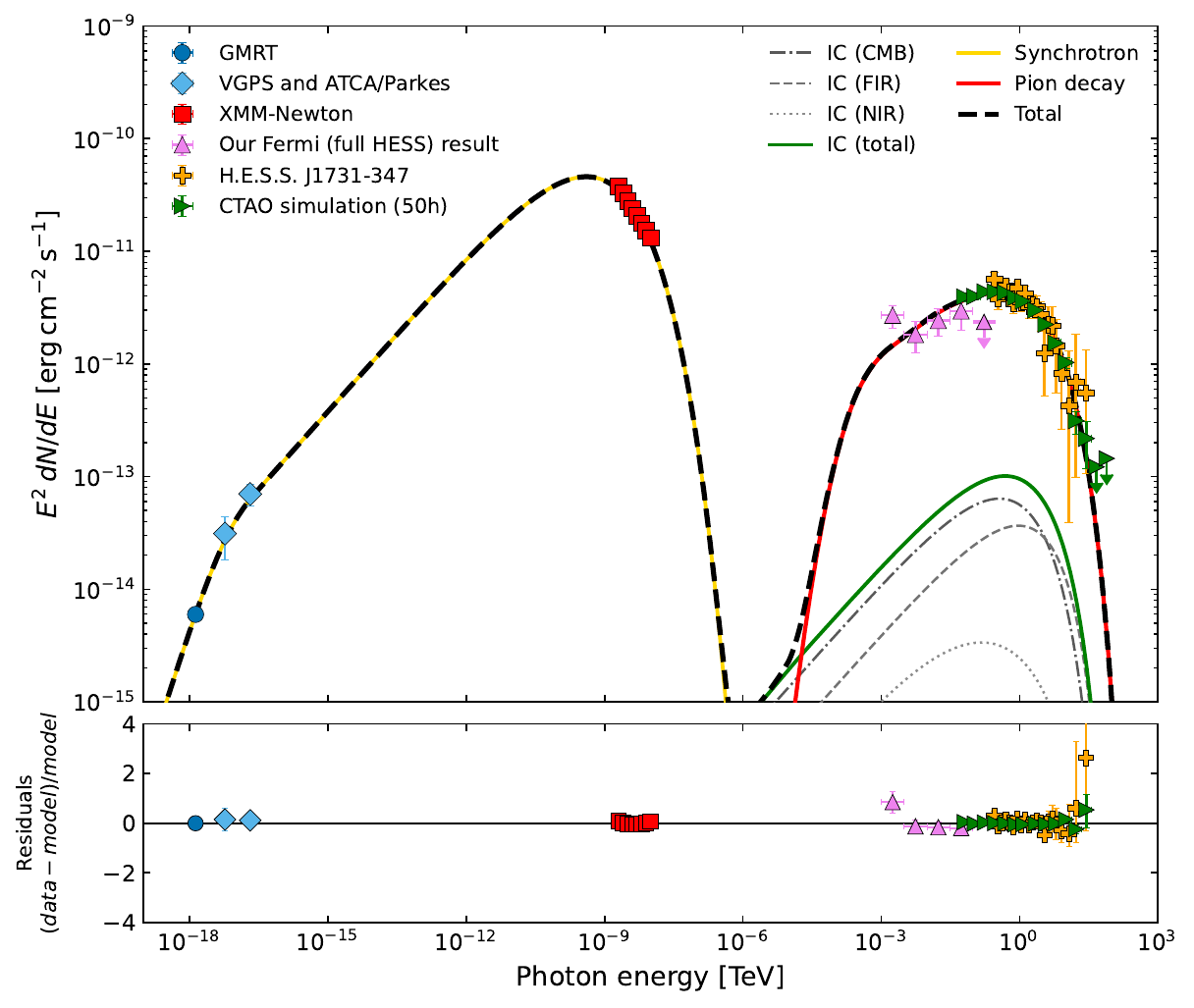}\\[0.3em]
    {\small (d) Lepto-hadronic scenario (full H.E.S.S. template)}
\end{minipage}
\caption{Broadband spectral energy distributions of HESS~J1731$-$347
including the H.E.S.S. measurements and the simulated CTAO flux points
(50~h, southern array, $z=20^\circ$) for the two emission scenarios. Panels (a) and (b) correspond
to the fiducial MC-core-excised configuration, with the rescaled H.E.S.S.
spectrum, and panels (c) and (d) to the integrated full-template
configuration, with the published H.E.S.S. points. The lower panels show the
fractional residuals of all datasets.}
\label{fig:sed_ctao}
\end{figure*}

Figure~\ref{fig:sed_ctao} places the simulated CTAO flux points on the
broadband SEDs of both configurations. With the 50~h exposure considered
here, the improved sensitivity and energy coverage of CTAO yield TeV flux
points with substantially smaller uncertainties than the current H.E.S.S.
measurements. We stress that these
simulations are based on the best-fit spectra of each scenario and do not
propagate the full posterior parameter space, so they illustrate the
discriminating power of CTAO for these specific models and do not constitute
a full forecast of the model comparison. Longer exposures of 200 to 300~h
could further tighten these constraints and allow spatially resolved spectral
analyses across the shell, with direct information on CR
acceleration and escape in this remnant.

% ============================================================
\section{Discussion}
\label{sec:discussion}
% ============================================================

The broadband analysis presented here revisits the origin of the
non-thermal emission from HESS~J1731$-$347, with particular attention
to the spatial definition of the GeV and TeV spectrum. Previous
\textit{Fermi}-LAT studies established a hard GeV counterpart associated with
the remnant \citep{Condon_2017,Guo_2018}, and \citet{Cui2019} reported a more
complex picture in which the emission toward the western MC-core/S0 region
is spectrally softer than that of the remaining SNR. With approximately
16~yr of LAT data, we recover the same qualitative tendency. S0 is nominally
softer than the MC-core-excised SNR component, although the difference
between the photon indices is only at the $\sim1.2\sigma$ level. We
therefore do not take the LAT result alone as evidence for distinct particle
populations, and use the spatial decomposition instead to test how the
treatment of the MC-core/S0 region propagates into the broadband modeling.

This test matters for a one-zone interpretation, because combining
$\gamma$-ray measurements from different spatial regions can bias the
inferred particle distributions. Within each of the two configurations, the
LAT and H.E.S.S. inputs are defined as consistently as possible, although the
MC-core-excised H.E.S.S. spectrum remains an energy-independent rescaling of
the published integrated spectrum and does not replace a spatially resolved
TeV spectral extraction. With this caveat, the information-criterion
comparison gives the same ordering in both configurations, with
$\Delta{\rm BIC}_{\rm noMC}=33.67$ for the fiducial dataset and
$\Delta{\rm BIC}_{\rm full}=37.95$ for the full-template dataset. Both values
indicate very strong evidence for the lepto-hadronic model on the scale of
\citet{Kass1995}. The relevant result is not the comparison between the two
$\Delta{\rm BIC}$ values, which refer to different datasets, but the fact that
the lower BIC of the lepto-hadronic model persists when the S0/MC-core region
is excluded from the SNR spectrum.

The two radiative interpretations imply very different partitions of the
non-thermal emission. In the fiducial leptonic solution, the same electron
population must reproduce the radio-to-X-ray synchrotron emission and the
$\gamma$-ray spectrum through IC scattering. The relatively low field,
$B=15.1\pm0.7~\mu$G, requires a large electron population, with
$W_e(E>1~{\rm GeV})= (2.15^{+0.19}_{-0.18})\times10^{48}$~erg,
which enhances the IC contribution. In the lepto-hadronic
solution, the field increases to $B=84.9^{+27.5}_{-22.9}~\mu$G, the synchrotron 
emission is reproduced with a much smaller electron population, with 
$W_e=(2.25^{+1.04}_{-0.65})\times10^{46}$ erg, and the remaining $\gamma$-ray 
emission is partly supplied by $\pi^0$ decay from a hard proton population with
$\alpha_p=1.69\pm0.10$ and $E_{\rm cut,p}=23.5^{+10.1}_{-5.9}$~TeV. The total
proton energy above 1~GeV, $W_p = (4.37^{+0.64}_{-0.51})\times10^{48}$~erg, 
corresponds to approximately $0.4\%$ of a canonical explosion energy of 
$10^{51}$~erg under the adopted $n_{\rm H}=20~{\rm cm^{-3}}$. The required proton 
energy budget is therefore modest, but its absolute value scales inversely with 
the assumed target density and cannot be interpreted independently of that 
assumption.

These results can be placed in the context of previous broadband studies.
The original H.E.S.S. analysis considered both IC and hadronic
interpretations without identifying the dominant $\gamma$-ray production
mechanism \citep{HESS2011}. After the detection of a hard GeV counterpart,
\citet{Guo_2018} showed that both purely leptonic and lepto-hadronic models
could reproduce the broadband SED, although they argued that the hard GeV
spectrum was more naturally accommodated by a leptonic interpretation within
standard DSA. The broadband datasets used by \citet{Guo_2018} and 
\citet{Cui2019} are largely similar to ours
in the radio, X-ray, and TeV bands, with the main
differences arising from the Fermi-LAT analyses
and the treatment of the spatially complex
GeV emission. A predominantly leptonic interpretation of the main SNR was
also adopted by \citet{Cui2016}, who invoked hadronic emission from escaped
cosmic rays to account for the nearby source HESS~J1729$-$345. More 
recently, \citet{Cui2019} separated the softer GeV component of the MC-core/
S0 region from the harder emission of the remaining SNR, describing the 
latter with a one-zone leptonic model and the former with hadronic 
interactions in a shock-cloud collision scenario.

Our analysis recovers the spatial trend found by \citet{Cui2019}, a
softer GeV component toward S0, but leads to a different model ordering for
the main remnant. In the fiducial configuration, where the MC-core region is
excluded from both the GeV and the rescaled TeV data, the lepto-hadronic
model still yields a substantially lower BIC than the purely leptonic model.
Within the one-zone assumptions adopted here, the statistical preference for
a hadronic contribution is therefore not driven solely by the spatially
distinct MC-core/S0 emission. This is the principal new result of the
present broadband analysis. The physical parameters of our lepto-hadronic
solution are nevertheless close to the hybrid solutions of \citet{Guo_2018},
with $B\simeq85~\mu$G and $\alpha_p\simeq1.69$ compared with
$B\simeq80$ to $87~\mu$G and $\alpha_p\simeq1.5$ to $1.7$. Fields of this
strength are compatible with magnetic-field amplification at young SNR
shocks \citep{Reynolds2012,Vink2012}. The hard proton spectrum, however,
remains a requirement imposed by the broadband data and not independent
evidence for standard DSA, since indices well below $\alpha\simeq2$ are not
trivial to produce in idealized strong-shock acceleration.

Several limitations should be kept in mind. The distance is fixed at
$d=3.2$~kpc, and the ambient density in the lepto-hadronic calculation is
fixed at $n_{\rm H}=20~{\rm cm^{-3}}$ and not inferred from the data.
HESS~J1731$-$347 evolves in a spatially inhomogeneous environment
\citep{Fukuda2014,Maxted2018}, so a single density and magnetic field cannot
represent the full shell. The MC-core-excised H.E.S.S. spectrum relies on an
energy-independent spatial rescaling. Radiative emission from secondary
electrons and positrons produced in charged-pion decays, including their IC
contribution, is not modeled explicitly. For the fiducial field of
$B\simeq85~\mu$G, the magnetic energy density is
$U_B=B^2/8\pi\simeq180~{\rm eV\,cm^{-3}}$, far above the total energy
density of the adopted photon fields, $U_{\rm rad}\simeq1.24~{\rm
eV\,cm^{-3}}$ including the CMB. Secondary pairs would therefore lose their
energy mainly through synchrotron radiation, which suggests that their IC
emission is suppressed. Their contribution has not been calculated
explicitly, however, and remains an uncertainty in the inferred particle
normalizations and energy budgets. Finally, the BIC
comparison should not be read as proof of a hadronic origin for
HESS~J1731$-$347. It quantifies the relative performance of the two specific
models considered here, penalizing the additional parameters of the
lepto-hadronic scenario, and does not exhaust the broader space of leptonic,
hadronic, or spatially resolved multi-zone descriptions.

The CTAO simulations of Section~\ref{sec:ctao} show how future data can
sharpen this comparison. For the best-fit models considered here, the two
spectra differ in curvature between a few hundred GeV and a few TeV, where a
50~h observation would yield the most precise flux points, and in their
decline at tens of TeV, where the lepto-hadronic spectrum falls more steeply
because $\pi^0$-decay photons carry only about one tenth of the parent proton
energy. These differences hold for one realization of each model. As shown
by RX~J1713.7$-$3946 \citep{HESS2018RXJ1713}, the TeV spectrum alone cannot
be expected to identify the radiating particles, and a quantitative
assessment of the discriminating power of CTAO would require propagating the
full posterior distributions of both models through the simulations and
fitting the competing models to the simulated data, which is beyond the scope
of this work.

% ============================================================
\section{Conclusions}
\label{sec:conclusions}
% ============================================================

The combination of a new 16~yr \textit{Fermi}-LAT analysis with Bayesian
broadband modeling shows that the inferred particle properties of
HESS~J1731$-$347 are sensitive to the spatial treatment of its complex GeV
emission, while the principal statistical result is preserved under both
spatial configurations. The treatment of the MC-core/S0 region produces only
small changes in the inferred electron population and more noticeable, but
statistically compatible, changes in the proton parameters. In both the
integrated and the MC-core-excised datasets, the lepto-hadronic model yields
a substantially lower BIC than the purely leptonic model.

For the fiducial MC-core-excised configuration, the lepto-hadronic
solution requires a magnetic field of $84.9^{+27.5}_{-22.9}~\mu$G, 
a hard proton spectrum with $\alpha_p = 1.69\pm 0.10$, and a proton energy content 
of $W_p = (4.37^{+0.64}_{-0.51})\times10^{48}$~erg for the adopted
$n_{\rm H}=20~{\rm cm^{-3}}$. These properties are broadly consistent with
previous hybrid modeling of the source, while the hard proton index and the
simplified one-zone description remain important limitations. Within the
models considered here, our results favor a non-negligible hadronic
contribution to the GeV and TeV emission, without excluding more complex
leptonic or multi-zone alternatives.

Our CTAO simulations show that a 50~h observation with the southern array
would detect HESS~J1731$-$347 at about $13\sigma$ to $18\sigma$, depending on the emission scenario and spatial configuration, and would resolve its shell morphology. For the specific best-fit models, the simulated spectra differ in curvature below a few TeV, where the CTAO flux points are most precise, and the lepto-hadronic spectrum steepens above $\sim5$~TeV while the leptonic spectrum keeps a harder tail. CTAO observations will therefore constrain the TeV spectral shape far better than current data and allow these specific broadband predictions to be tested. Because the simulations are based on the best-fit spectra and do not explore the full posterior parameter space, they do not establish that CTAO can distinguish every viable leptonic and hadronic interpretation of the source. Longer exposures of 200 to 300~h would further tighten the constraints and could reveal spatial variations of the $\gamma$-ray spectrum across the shell, with direct consequences for our understanding of CR acceleration in young supernova remnants.

\begin{acknowledgments}
We thank the anonymous referee for their valuable suggestions and comments. D.B.G. acknowledges the fellowship from Fundação Araucária, Brazil (No. 501/2024). The authors thank Rubens Jr. Costa for insightful comments on CTAO simulations. M.F.S. thanks the Fundação de Amparo à Pesquisa do Estado de São Paulo (FAPESP, grants No. 2026/06733-0, No. 2025/05794-2 and No. 2021/01089-1) for financial support. The research of R.C.A. is supported by the CAPES/Alexander von Humboldt Program (88881.800216/2022-01), CNPq (308859/2025-1, 4000045/2023-0), Araucária Foundation (698/2022, 721/2022) and FAPESP (2021/01089-1). D.B.G. and R.C.A. also acknowledge the support of the NAPI “Fenômenos Extremos do Universo” of Fundação de Apoio à Ciência, Tecnologia e Inovação do Paraná.

\noindent This research has made use of the CTA instrument response functions provided by the CTA Consortium and Observatory, see \url{https://www.ctao-observatory.org/science/cta-performance/} (version prod6 v1.0, \cite{cta_observatory_2026_22871179}) for more details. This work also made use of Gammapy \citep{2023A&A...678A.157D}, a community-developed Python package. The Gammapy team acknowledges all past and current contributors to Gammapy, as well as all contributors of the main Gammapy dependency libraries, namely \href{https://numpy.org/}{NumPy}, \href{https://scipy.org/}{SciPy}, \href{http://www.astropy.org}{Astropy}, \href{https://astropy-regions.readthedocs.io/}{Astropy Regions}, \href{https://scikit-hep.org/iminuit/}{iminuit}, and \href{https://matplotlib.org/}{Matplotlib}.

\noindent ChatGPT (OpenAI) was used to assist with language editing and
refinement of the manuscript text. All scientific content and
interpretations were developed and verified by the authors, who take
full responsibility for the final manuscript.
\end{acknowledgments}

\software{\texttt{naima}~\citep{Zabalza2015}, \texttt{Gammapy}~\citep{2023A&A...678A.157D, gammapy}, \texttt{Astropy}~\citep{astropy_2022}, \texttt{Fermipy}~\citep{Wood2017}, \texttt{Fermitools}~\citep{Fermitools2019}, \texttt{emcee}~\citep{Foreman-Mackey2013}}

\bibliography{references}{}
\bibliographystyle{aasjournalv7.1}

\end{document}